\pdfoutput=1
\documentclass[aps, twocolumn, floatfix, superscriptaddress, prl]{revtex4-1}
\usepackage{graphicx}
\DeclareGraphicsExtensions{.pdf}
\usepackage{amsmath,amssymb,bbold,bm,color}
\usepackage{float}
\usepackage{epstopdf}
\usepackage{tabularx}
\usepackage{braket}
\usepackage{booktabs}
\usepackage{array}
\usepackage{blindtext}
\usepackage{gensymb}
\usepackage{dutchcal}
\usepackage[colorlinks=true,breaklinks=true,linkcolor=blue,anchorcolor=red,citecolor=blue,urlcolor=blue]{hyperref}
\usepackage{seqsplit}
\allowdisplaybreaks[1]

\newcommand{\sgn}{{\text{sgn}}}

\begin{document}

\title{Chiral Anomaly Beyond Fermionic Paradigm}

\author{Tianyu Liu}
%\email{tliu@pks.mpg.de}
\affiliation{Shenzhen Institute for Quantum Science and Engineering and Department of Physics, Southern University of Science and Technology (SUSTech), Shenzhen 518055, China}
\affiliation{International Quantum Academy, Shenzhen 518048, China}
\affiliation{Max-Planck-Institut f\"ur Physik komplexer Systeme, 01187 Dresden, Germany}

\author{Zheng Shi}
%\email{zheng.shi@uwaterloo.ca}
\affiliation{Institute for Quantum Computing and Department of Physics and Astronomy, University of Waterloo, Waterloo, Ontario, N2L 3G1, Canada}
\affiliation{Dahlem Center for Complex Quantum Systems and Physics Department, Freie Universit\"at Berlin, 14195 Berlin, Germany}

\author{Hai-Zhou Lu}
\email{Corresponding author: luhz@sustech.edu.cn}
%\email{luhz@sustech.edu.cn}
\affiliation{Shenzhen Institute for Quantum Science and Engineering and Department of Physics, Southern University of Science and Technology (SUSTech), Shenzhen 518055, China}
\affiliation{Quantum Science Center of Guangdong-Hong Kong-Macao Greater Bay Area (Guangdong), Shenzhen 518045, China}
\affiliation{Shenzhen Key Laboratory of Quantum Science and Engineering, Shenzhen 518055, China}
\affiliation{International Quantum Academy, Shenzhen 518048, China}

\author{X. C. Xie}
\affiliation{International Center for Quantum Materials, School of Physics, Peking University, Beijing100871, China}
\affiliation{Institute for Nanoelectronic Devices and Quantum Computing, Fudan University, Shanghai 200433, China}
\affiliation{Hefei National Laboratory, Hefei 230088, China}

\begin{abstract} 
Two-dimensional magnets have manifested themselves as promising candidates for quantum devices. We here report that the edge and strain effects during the device fabrication with two-dimensional honeycomb ferromagnets such as CrX$_3$ (X=Cl, I, Br) and CrXTe$_3$ (X=Si, Ge) can be characterized by a (1+1)-dimensional magnon chiral anomaly beyond the fermionic paradigm. In the presence of zigzag edges, a pair of chiral bulk-edge magnon bands appear and cause an imbalance of left- and right-chirality magnons when subjected to nonuniform temperature or magnetic fields. In the presence of a uniaxial strain, the bulk Dirac magnons are broken into chiral magnon pseudo-Landau levels, resulting in a magnon chiral anomaly observable through a negative strain-resistivity of the magnetic dipole and heat. Our work demonstrates a chiral anomaly with (quasi)particles obeying non-fermionic statistics and will be instructive in understanding anomalous magnon transport. 
\end{abstract}

\date{\today}
\maketitle

\textit{Introduction.---} Two-dimensional magnets recently attract extensive attention in the context of quantum information, quantum computation, and spintronics devices~\cite{burch2018, chumak2015,  yuan2022, chumak2012, chumak2014, vogt2014}. One prominent type of such materials includes honeycomb ferromagnets~\cite{chenlebing2022, caizhengwei2021, pershoguba2018, kim2019, chenlebing2018, huang2017, chenlebing2021, zhufengfeng2021}, whose magnonic band structures exhibit Dirac cones resembling the electronic band structures of graphene~\cite{castroneto2009, novoselov2004} and Weyl/Dirac semimetals~\cite{armitage2018, wan2011, burkov2011, young2012, wang2012, wang2013}. Such resemblance renders honeycomb ferromagnets promising candidates for topological spintronics devices, which can support non-dissipative spin transport~\cite{chenlebing2018, pershoguba2018}. The fabrication of topological spintronics devices inevitably introduces edges and strain~\cite{huang2017, yu2022}, which can drastically alter the properties of the devices. Therefore, it is crucial to understand the edge and strain effects in a quantitative fashion.

\begin{figure} [t!]
\includegraphics[width = 8.6cm]{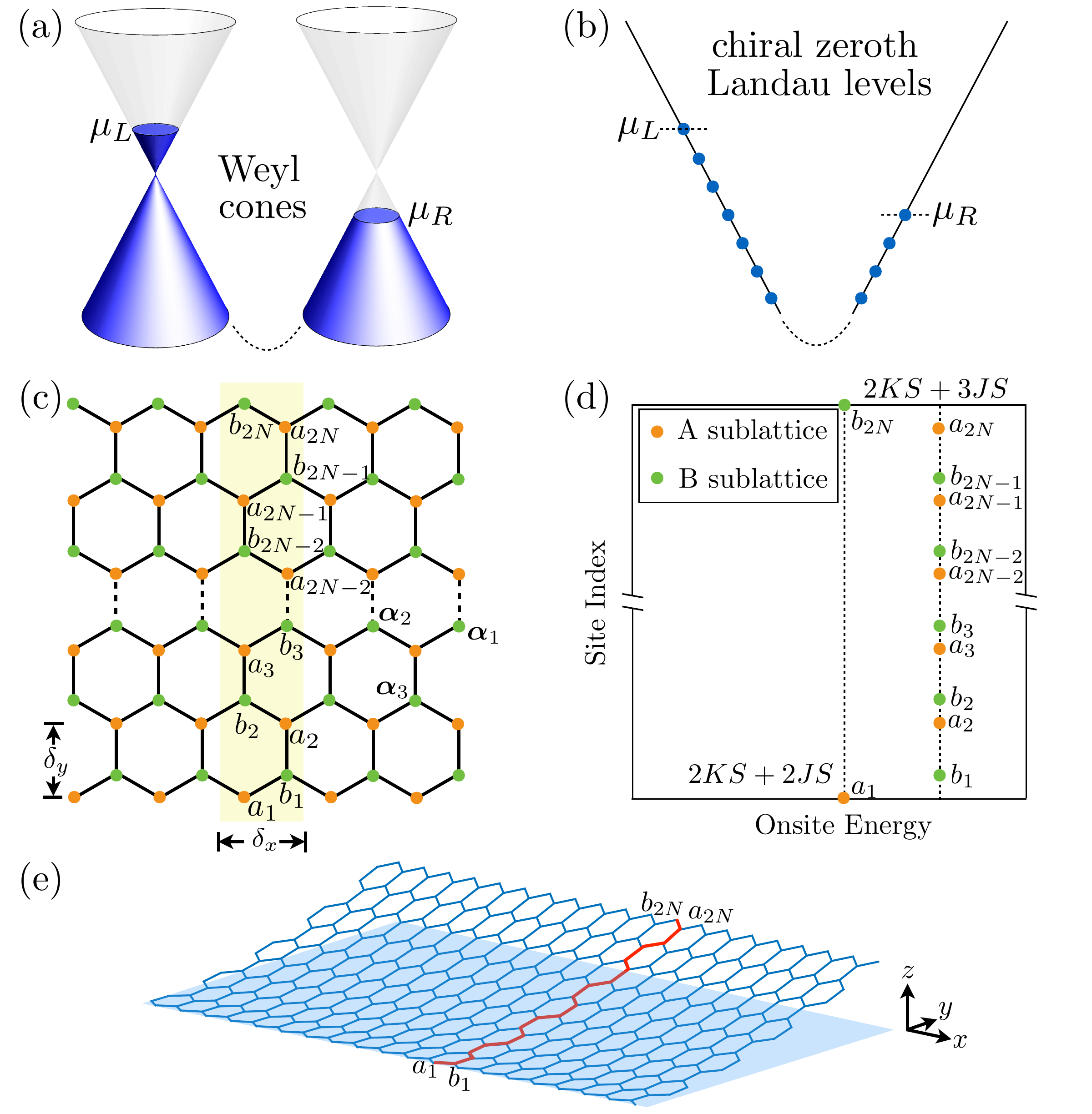}
\caption{Schematics of the chiral anomaly of Weyl semimetals in the (a) semiclassical and (b)  quantum limits. (c) Zigzag ribbon of a honeycomb ferromagnet. The yellow shade highlights its unit cell of width $\delta_x$ and length $W=2N\delta_y$. (d) Onsite energy of each spin-$S$ lattice site of the unit cell with $J$ being the nearest-neighbor interaction and $K$ measuring the out-of-plane easy-axis anisotropy. (e) A strained zigzag ribbon with the blue shade marking its position without strain.
} \label{fig1}
\end{figure}

\begin{figure} [t]
\includegraphics[width = 8.6cm]{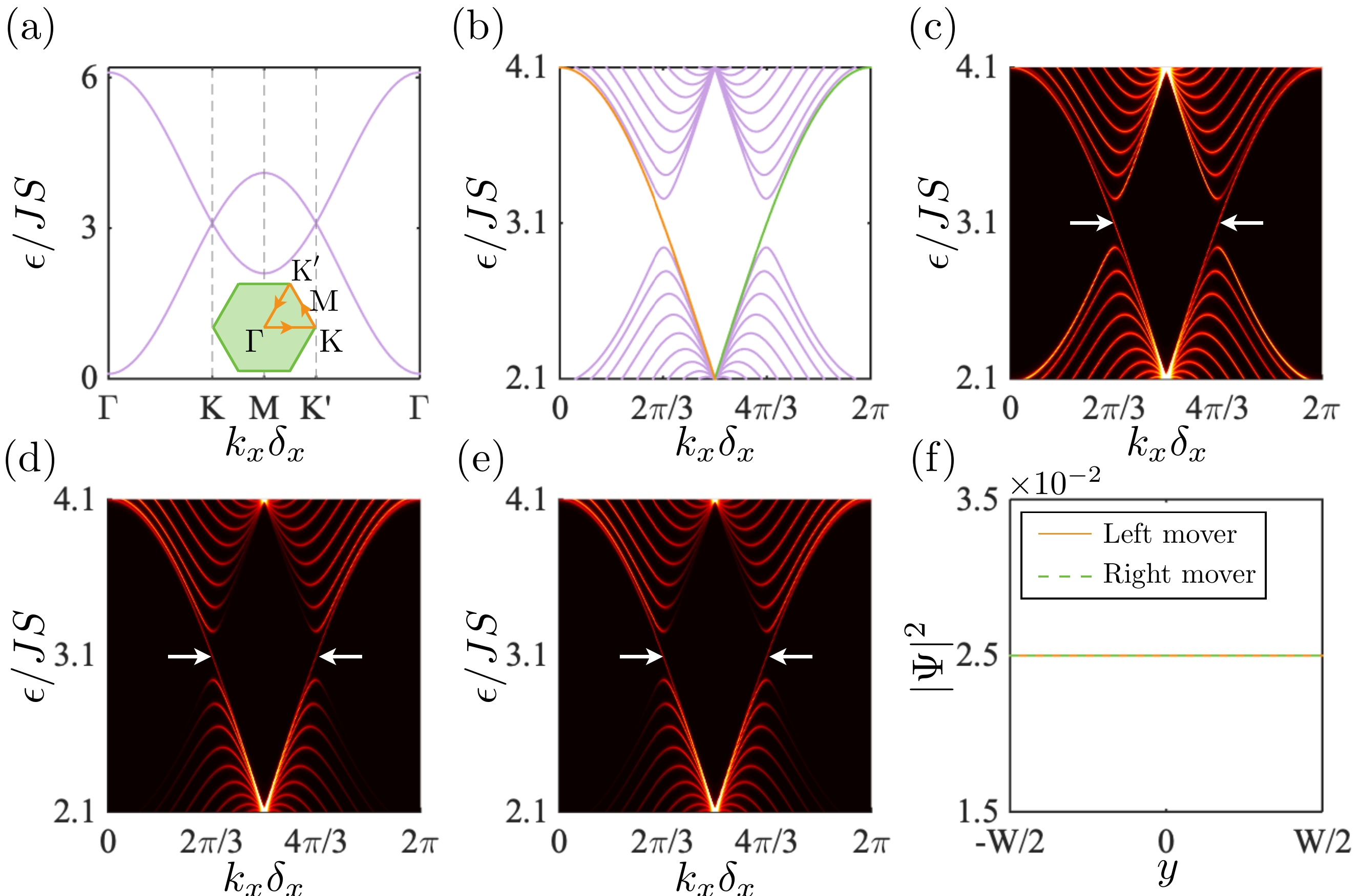}
\caption{(a) Magnon bands of a honeycomb ferromagnet exhibit Dirac cones at Brillouin zone corners. (b) Magnon bands (purple) of a zigzag ribbon of a honeycomb ferromagnet. The analytical dispersions [Eq.~(\ref{bulk-edge})] of the right-moving (green) and left-moving (orange) chiral magnon bands are overlaid. (c)--(e) are respectively the magnon spectral functions of the bulk, top edge, and bottom edge. The white arrows highlight the chiral magnon bands. (f) Magnon probability distributions of the right-moving (green dashed) and left-moving (orange solid) chiral magnon bands computed at $k_x\delta_x=\tfrac{4\pi}{3}$ and $k_x\delta_x=\tfrac{2\pi}{3}$ respectively. We set $K=0.05J$ for all panels.
} \label{fig2}
\end{figure}

\begin{figure} [t]
\includegraphics[width = 8.6cm]{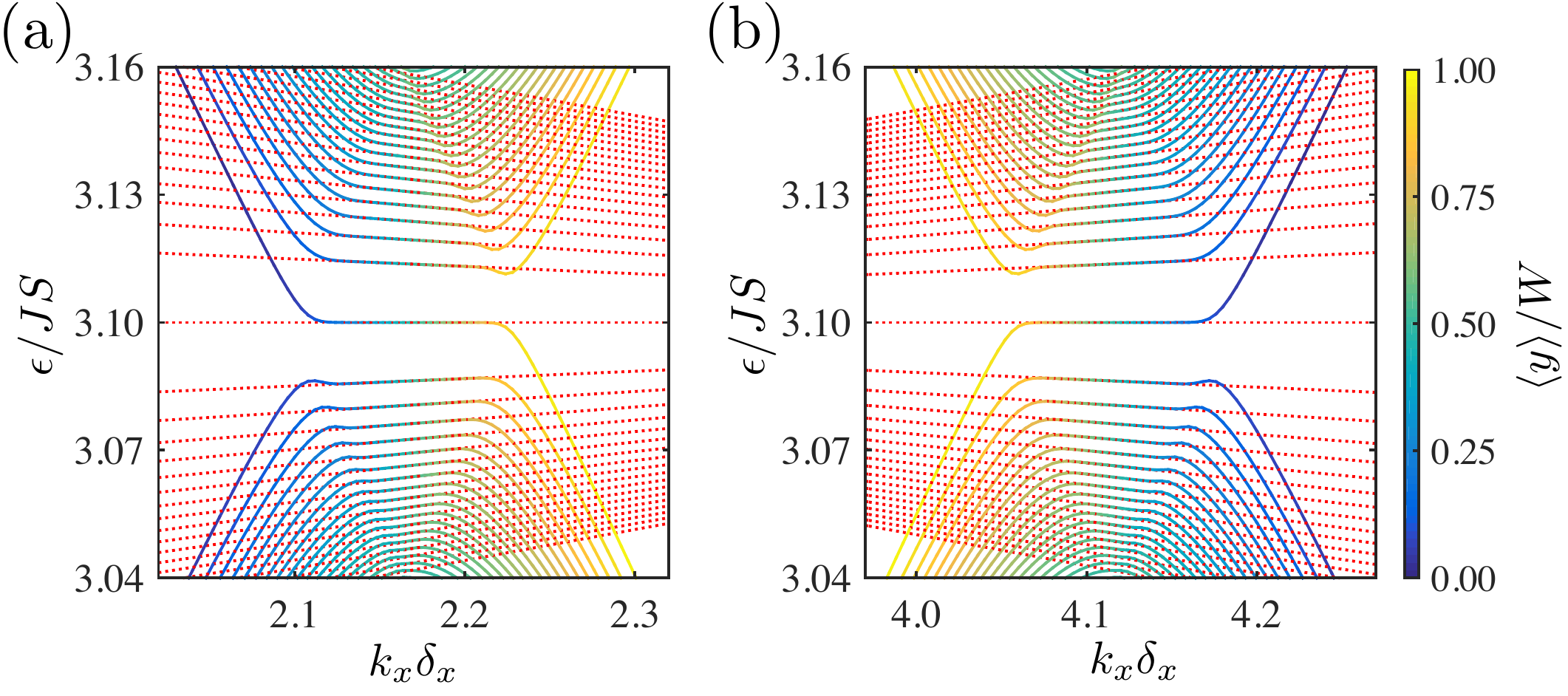}
\caption{Chiral magnon pseudo-Landau levels around the projected Brillouin zone corners (a) $K$ and (b) $K'$ [Inset, Fig.~\ref{fig2}(a)]. The curves are colored by the expectation of the position operator, i.e., $\braket y=\bra{\Psi_{k_x}(y)} y \ket{\Psi_{k_x}(y)}$. The analytical dispersions [Eq.~(\ref{pLL})] of magnon pseudo-Landau levels are overlaid as dotted curves. We set the easy-axis anisotropy $K=0.05J$ and the strain $\gamma \lambda \delta_x = 1.73 \times 10^{-4}$. 
} \label{fig3}
\end{figure}

\begin{figure} [t]
\includegraphics[width = 8.6cm]{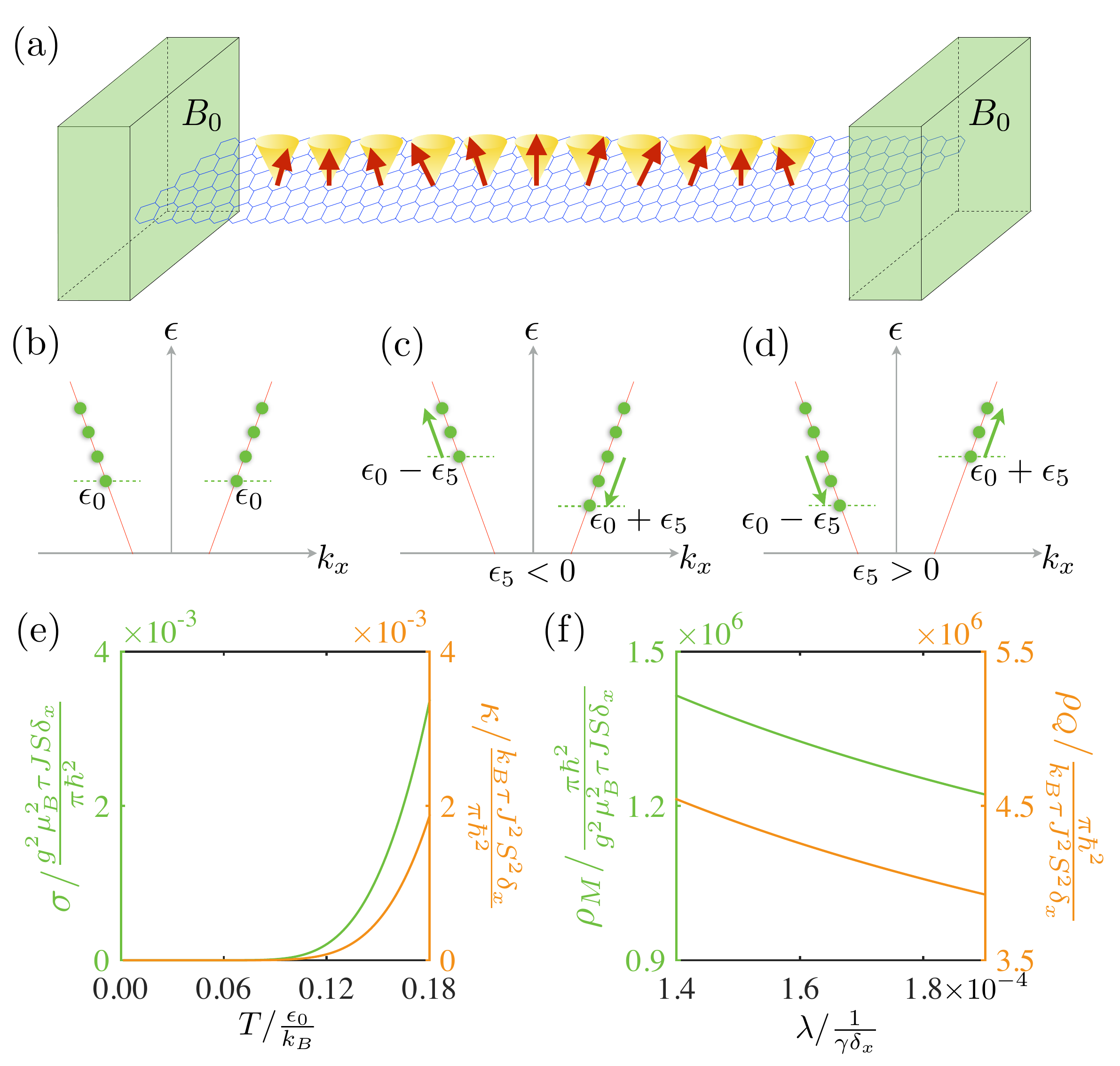}
\caption{Magnon transport. (a) Proposed experimental apparatus comprises two magnon reservoirs (green) connected to the ends of the zigzag ribbon (blue), which carries the spin wave (cones and arrows). (b) Magnon population of the chiral bands in the absence of driving forces, with energy bounded by $\epsilon_0$ from below. (c, d) Magnon populations when reservoirs are driven out of equilibrium. (e) Dipole conductivity (green) and thermal conductivity (orange) associated with the chiral bulk-edge magnon bands of the strain-free ribbon. We place the population edge at $\epsilon_0=3.1JS$. (f) Dipole resistivity (green) and thermal resistivity (orange) associated with the first magnon pseudo-Landau level in the uniaxially strained ribbon. We set $\epsilon_0=3.1132JS$ and $T=0.1\epsilon_0/k_B$.
} \label{fig4}
\end{figure}

In this Letter, we show that the edge and strain effects of honeycomb ferromagnets can be understood in terms of chiral anomaly, the non-conservation of the chiral charge because of quantization~\cite{adler1969, bell1969, nielsen1983}. For Weyl semimetals, the electronic chiral anomaly~\cite{huang2015, kim2013, xiong2015, son2013, zhang2016, burkov2015, liqiang2016, fukushima2008} arises from the intervalley imbalance of chiral charge [Fig.~\ref{fig1}(a)] mediated by the chiral zeroth Landau levels [Fig.~\ref{fig1}(b)]. Remarkably, similar chiral bands can emerge in the magnon spectrum of honeycomb ferromagnets when edges are created or strain is applied, and support a magnon chiral anomaly beyond the fermionic paradigm. On the one hand, the creation of edges modulates the onsite magnon energy, which is related to the number of nearest neighbors a site possesses. For a zigzag ribbon of honeycomb ferromagnets [Fig.~\ref{fig1}(c)], we show that the modulation [Fig.~\ref{fig1}(d)] induced by the edge creation results in chiral bulk-edge magnon bands [Fig.~\ref{fig2}(b)], which resemble the chiral zeroth Landau levels in Weyl semimetals [Fig.~\ref{fig1}(b)] and can produce a (1+1)-dimensional magnon chiral anomaly in the presence of nonuniform temperature or magnetic fields. On the other hand, as strain is famously known to act as a gauge field inducing pseudo-Landau levels in Dirac matter~\cite{rachel2016, perreault2017, liu2017b, matsushita2018, massarelli2017, nica2018, rechtsman2013, wen2019, brendel2017, peri2019, ferreiros2018, sunjunsong2021, sunjunsong2022, guinea2010a, levy2010, liu2017a, liu2020a, pikulin2016, grushin2016}, the Dirac magnons of honeycomb ferromagnets [Fig.~\ref{fig2}(a)] should be Landau-quantized by strain. We show that a tensile uniaxial strain $\epsilon_{yy}=\lambda y$ generated by an out-of-plane lattice deformation [Fig.~\ref{fig1}(e)] can produce, with a reasonably small $\lambda\sim10^{-4}\,\text{nm}^{-1}$, chiral magnon pseudo-Landau levels [Figs.~\ref{fig3}(a) and~\ref{fig3}(b)], which also yield a (1+1)-dimensional magnon chiral anomaly. The dispersion of such magnon pseudo-Landau levels is attributed to the strain-modulated Dirac cone velocity~\cite{liu2021a}. The transport associated with the (1+1)-dimensional magnon chiral anomaly is investigated using the Boltzmann formalism. In particular, for the magnon chiral anomaly mediated by the chiral pseudo-Landau levels, both magnetic dipole and heat transport exhibit a negative strain-resistivity [Fig.~\ref{fig4}(f)] analogous to the negative magnetoresistivity in Weyl semimetals~\cite{huang2015, kim2013, xiong2015, son2013, zhang2016, burkov2015, liqiang2016, fukushima2008}. We propose that the (1+1)-dimensional magnon chiral anomaly may be experimentally observable in zigzag ribbons of van der Waals ferromagnets such as CrX$_3$ (X=Cl, I, Br)~\cite{chenlebing2022, caizhengwei2021, chenlebing2018, huang2017, chenlebing2021, kim2019, pershoguba2018} and CrXTe$_3$ (X=Si, Ge)~\cite{zhufengfeng2021}.

\textit{Model.---} The honeycomb ferromagnets in consideration can be characterized by a simple Heisenberg Hamiltonian
\begin{equation} \label{Heisenberg}
H=- \sum_{\langle i,j \rangle}J_{ij} \bm S_i\cdot\bm S_j - \sum_{i} K_i(\bm S_i\cdot\hat z)^2,
\end{equation} 
where $\bm S_i$ is the spin on the $i$th lattice site, the first term represents the nearest-neighbor ferromagnetic exchange interaction with $J_{ij}>0$, and the second term indicates a $z$ direction easy axis with $K_i>0$. Such an out-of-plane easy axis is favored in honeycomb ferromagnets CrBr$_3$~\cite{caizhengwei2021, pershoguba2018, kim2019}, CrI$_3$~\cite{kim2019, chenlebing2018, huang2017, chenlebing2021} and CrXTe$_3$ (X=Si, Ge)~\cite{zhufengfeng2021}, where the nearest-neighbor interaction $J_{ij}$, the anisotropy $K_i$, and the spin magnitude $S_i$ do not depend on lattice site positions. We set in the following $J_{ij}\equiv J$, $K_i\equiv K$, and $S_i\equiv S$ unless otherwise specified.

To describe magnons of the Heisenberg Hamiltonian [Eq.~(\ref{Heisenberg})] we perform the Holstein-Primakoff transformation~\cite{holstein1940} $S_i^+= (2S-\mu_i^\dagger \mu_i)^{1/2} \mu_i= (S_i^-)^\dagger $ and $S_i^z = S-\mu_i^\dagger \mu_i$, where $\mu_i^\dagger$ ($\mu_i$) is the magnon creation (annihilation) operator on the $i$th lattice site belonging in the $\mu$th sublattice ($\mu=a,b$). In the large-$S$ limit, the magnon-magnon interactions can be ignored, leading to a magnon tight-binding Hamiltonian
\begin{align} \label{TB}
H = \mathcal K \sum_{i} (a_i^\dagger a_i +b_i^\dagger b_i) -JS  \sum_{\langle i,j \rangle} ( a_i^\dagger b_j + b_j^\dagger a_i ),
\end{align}
where $\mathcal K=2KS+3JS$ is the magnon onsite energy. Equation~(\ref{TB}) exhibits magnon Dirac cones at the corners of the hexagonal Brillouin zone [Fig.~\ref{fig2}(a)], in close resemblance to the nearest-neighbor tight-binding model of graphene. Consequently, the confinement-induced chiral bands~\cite{shen2017, lichangan2019} and strain-induced chiral pseudo-Landau levels~\cite{lantagne2020, shi2021, liu2022} of graphene should also find their magnonic analogs in honeycomb ferromagnets.

\textit{Chiral bulk-edge magnon bands.---} The ribbon of honeycomb ferromagnets defines the (1+1)-dimensional spacetime in consideration. For a zigzag ribbon along the $x$ direction [Fig.~\ref{fig1}(c)], a slight modification of Eq.~(\ref{TB}) followed by a partial Fourier transform leads to a Bloch Hamiltonian~\cite{suppl}
\begin{equation} \label{Bloch}
\mathcal H_{k_x} = 
\renewcommand*{\arraystretch}{1.5} 2KS+ JS
\begin{pmatrix}
\bm\Lambda_A & -2\cos(\tfrac{1}{2}k_x\delta_x)\bm I-\bm L
\\
* & \bm\Lambda_B
\end{pmatrix},
\end{equation}
where we define for transparency four $2N\times2N$ matrices: $\bm I$ is the identity matrix, $\bm L$ has nonzero entries $L_{n,n-1}=1$ for $n=2,\cdots,2N$, and $\bm \Lambda_A=\text{diag}(2,3,\cdots,3,3)$ and $\bm \Lambda_B=\text{diag}(3,3,\cdots,3,2)$ record the numbers of nearest neighbors for each lattice site. Reflected in $\bm \Lambda_{A,B}$, the edge sites experience an onsite energy $2KS+2JS=\mathcal K-JS$ different from that of the bulk sites $\mathcal K$ [Fig.~\ref{fig1}(d)]. Such modulation of the magnon onsite energy is analogous to the confinement in graphene~\cite{shen2017, lichangan2019}, and thus results in chiral magnon bands as expected [Fig.~\ref{fig2}(b)]. Interestingly, these chiral magnon bands exhibit nonzero spectral weights both in the bulk [Fig.~\ref{fig2}(c)] and on the zigzag edges [Figs.~\ref{fig2}(d) and~\ref{fig2}(e)], revealing their bulk-edge nature. Solving the Harper's equations associated with $\mathcal H_{k_x}$, we find that the right-moving and left-moving chiral bulk-edge magnon bands disperse according to~\cite{suppl}
\begin{equation} \label{bulk-edge}
\epsilon_{k_x,R/L} =2KS+ 2JS \mp 2JS \cos \left(\frac{1}{2} k_x\delta_x\right).
%\begin{split}
%\epsilon_{k_x,R} &= 2KS+2JS - 2JS \cos \left(\frac{1}{2} k_x\delta_x\right), 
%\\
%\epsilon_{k_x,L} &=2KS+ 2JS + 2JS \cos \left(\frac{1}{2} k_x\delta_x\right).
%\end{split}
\end{equation}
Their eigenvectors are $\ket{\Psi_R}=\tfrac{1}{\sqrt{4N}}(\mathcal r,\mathcal r)^T$ and $\ket{\Psi_L}=\tfrac{1}{\sqrt{4N}}(\mathcal l,-\mathcal l)^T$, where $\mathcal r=(1,1,\cdots,1,1)$ and $\mathcal l=(1,-1,\cdots,1,-1)$ are $2N$-component vectors. The explicit forms of $\ket{\Psi_{R,L}}$ unambiguously show the spatially uniform distribution of the magnon probability [Fig.~\ref{fig2}(f)] associated with the chiral bands [Eq.~(\ref{bulk-edge})] and further evince their bulk-edge nature as reflected by the spectral functions [Figs.~\ref{fig2}(c)--\ref{fig2}(e)].%$\ket{\Psi_R}=\tfrac{1}{\sqrt{4N}}(\mathcal r,\mathcal r)^T$ and $\ket{\Psi_L}=\tfrac{1}{\sqrt{4N}}(\mathcal l,-\mathcal l)^T$

\textit{Chiral magnon pseudo-Landau levels.---} A proper strain in graphene acts as a $U(1)$ gauge field and modulates the Dirac cone velocity to induce chiral pseudo-Landau levels~\cite{lantagne2020, shi2021, liu2022}. The resemblance between honeycomb ferromagnets and graphene [Eq.~(\ref{TB})] thus suggests similar chiral pseudo-Landau levels of magnons in strained honeycomb ferromagnets. We here consider a uniaxial strain $\epsilon_{yy} = \lambda y$, known to induce chiral pseudo-Landau levels in graphene~\cite{lantagne2020}. Such a strain pattern may arise from an out-of-plane lattice deformation [Fig.~\ref{fig1}(e)] profiled by $h(x,y)=\sqrt{8\lambda y^3/9}$ in the half-plane $y\geq 0$~\cite{suppl}. By varying the bond lengths of the lattice, the strain modulates the nearest-neighbor interaction $J_{ij}$ in an approximately exponential fashion as
\begin{align} \label{J}
J_{ij}=J\exp \left\{ -\gamma \frac{[ (\bm r_j - \bm r_i)\cdot \bm \nabla h]^2}{2|\bm r_j - \bm r_i|^2} \right\},
\end{align}
where $\bm r_{i(j)}$ labels the position of the $i(j)$th lattice site without strain and $\gamma \simeq 1$ is the Gr\"uneisen parameter of the honeycomb ferromagnet in consideration. Depending purely on $y$, the profiled lattice deformation preserves the translational symmetry along the $x$ direction. The strained ribbon is then characterized by 
\begin{align} \label{TB_ribbon}
\tilde H&=\sum_{k_x,y}\bigg[ \mathcal K \left(a_{k_x,y}^\dagger a_{k_x,y} + b_{k_x,y}^\dagger b_{k_x,y} \right) -\bigg\{ b_{k_x,y}^\dagger\nonumber\\
\times& \left[ 2J_1(y)\cos\frac{k_x\delta_x}{2}+J_3(y) \hat s_{\delta_y} \right]S a_{k_x,y} +\mathrm{h.c.}\bigg\}\bigg] ,
\end{align}
%\tilde H&=\sum_{k_x,y}\mathcal K \left( a_{k_x,y}^\dagger a_{k_x,y} + b_{k_x,y}^\dagger b_{k_x,y} \right) \\
%&-\sum_{k_x,y}  b_{k_x,y}^\dagger \left[ 2J_1(y)\cos\left(\frac{1}{2}k_x\delta_x\right)+J_3(y) \hat s_{\delta_y} \right]S a_{k_x,y}  \nonumber\\
%&-\sum_{k_x,y}  a_{k_x,y}^\dagger \left[ 2J_1(y)\cos \left(\frac{1}{2}k_x\delta_x \right)+J_3(y) \hat s_{-\delta_y} \right]S b_{k_x,y},  \nonumber
%
where $J_1(y)=J\exp(-\tfrac{1}{4}\gamma\lambda y)$, $J_3(y)=J\exp(-\gamma\lambda y)$, and $\hat s_{\delta_y}$ shifts the annihilation operators according to $\hat s_{\delta_y}\mu_{k_x,y}=\mu_{k_x,y +\delta_y}$~\cite{suppl}. For simplicity, we ignore in Eq.~(\ref{TB_ribbon}) the strain-induced onsite energy, because it can be compensated by a Zeeman field~\cite{suppl}; we also ignore in analytics the edge terms in the onsite energy, which are unimportant in the following discussion. Note that $\tilde H$ at a given $k_x$ realizes a Su-Schrieffer-Heeger model~\cite{su1979}, whose intracell hopping $-2J_1(y)S \cos(\tfrac{1}{2}k_x\delta_x)$ and intercell hopping $-J_3(y)S$ both depend smoothly on $y$. The two hoppings become equal at
%, whose intracell (i.e., between $a_{k_x,y}$ and $b_{k_x,y}$) magnon hopping reads $-2J_1(y)S \cos(\tfrac{1}{2}k_x\delta_x)$ and intercell (i.e., between $b_{k_x,y}$ and $a_{k_x,y+\delta_y}$) magnon hopping is $-J_3(y)S$. A domain wall is then found at
%
\begin{equation} \label{y0}
y_0=-\frac{4}{3\gamma\lambda} \ln \left|-2\cos\left(\frac{1}{2}k_x\delta_x\right)\right|.
\end{equation}
%y_0=-\frac{4}{3\gamma\lambda} \ln \left|-2\cos\left(\frac{1}{2}k_x\delta_x\right)\right|,
%
The $y<y_0$ and $y>y_0$ sectors of the ribbon are dominated by intercell and intracell hoppings respectively. Therefore, a domain wall exists at the location $y=y_0$ and carries a topologically protected zero mode analogous to the zero-energy end modes of polyacetylene~\cite{su1979}, as long as it is inside the ribbon and not too close to an edge. The domain wall modes at different $k_x$'s thus constitute a flat bulk magnon band  [Figs.~\ref{fig3}(a) and \ref{fig3}(b)] at the energy $\mathcal K=2KS+3JS$. 

We now show that the flat bulk magnon band is by nature the strain-induced zeroth magnon pseudo-Landau level. Linearizing $\tilde H$ in the vicinity of the domain wall $y_0$, we obtain its associated Bloch Hamiltonian as~\cite{suppl}
\begin{equation} \label{Dirac}
\begin{split}
\tilde{\mathcal H}_{k_x,y}&= \frac{3}{4}JS\gamma\lambda\left|-2\cos\left(\frac{1}{2}k_x\delta_x\right)\right|^{\frac{4}{3}}(y-y_0)\sigma^x
\\
&+\frac{3}{2}iJSa \left|-2\cos\left(\frac{1}{2}k_x\delta_x\right)\right|^{\frac{4}{3}} \partial_y\sigma^y+\mathcal K \sigma^0,
\end{split}
\end{equation}
where $\sigma^{x,y}$ and $\sigma^0$ are respectively Pauli matrices and the identity matrix in sublattice space. This real-space linearization remarkably leads to an exactly solvable Dirac theory [Eq.~(\ref{Dirac})], whose spectrum comprises magnon pseudo-Landau levels
\begin{equation} \label{pLL}
\epsilon_{\nu,k_x}=\mathcal K+\frac{3}{2}\sgn(\nu)JS\sqrt{|\nu|\gamma\lambda a}\left|2\cos\left(\frac{1}{2}k_x\delta_x\right) \right|^{\frac{4}{3}}.
\end{equation}
In Eq.~(\ref{pLL}), $\nu$ is an integer and the zeroth pseudo-Landau level $\epsilon_{0,k_x}=\mathcal K=2KS+3JS$ indeed coincides with the aforementioned flat bulk magnon band. More interestingly, the non-zeroth pseudo-Landau levels $\epsilon_{\nu \neq 0,k_x}$ are all chiral, because strain modulates the Dirac cone velocity as reflected in Eq.~(\ref{Dirac}).

\textit{Magnon chiral anomaly.---} The chiral bulk-edge magnon bands [Eq.~(\ref{bulk-edge})] and chiral magnon pseudo-Landau levels [Eq.~(\ref{pLL})] in zigzag ribbons of honeycomb ferromagnets [Figs.~\ref{fig1}(c) and~\ref{fig1}(e)] are analogous to the chiral zeroth Landau levels in Weyl semimetals [Fig.~\ref{fig1}(b)], and thus provide the necessary band structure for the (1+1)-dimensional magnon chiral anomaly in consideration. To experimentally implement this magnon chiral anomaly, we adopt a two-end apparatus~\cite{meier2003} comprising two magnon reservoirs in contact with a zigzag ribbon [Fig.~\ref{fig4}(a)], which is shorter than the magnon mean free path $\sim100\,\text{nm}$~\cite{lisy2005, boona2014} to guarantee ballistic transport. The reservoirs are in thermal equilibrium with the ribbon at the temperature $T$ and subjected to a uniform magnetic field $B_0=\epsilon_0/g\mu_B$ with $\epsilon_0\simeq \mathcal K$ such that the magnons inside possess an energy no lower than $\epsilon_0$. When appropriately excited by microwaves, the magnons inside the left (right) reservoir are pumped to the right (left), tunnel into the ribbon, and start to occupy the right-moving (left-moving) chiral bands from $\epsilon_0$ [Fig.~\ref{fig4}(b)]. We refer to $\epsilon_0$ as the ``magnon population edge''  because it bounds magnons from below (cf., Fermi level bounding electrons from above).

When a magnetic field gradient $\partial_x B$ and a temperature gradient $\partial_x T$ are exerted on the ribbon by driving the reservoirs out of equilibrium, magnons begin to march along the chiral bands according to the semiclassical equation of motion~\cite{huebener2001}
\begin{equation} \label{eom}
\hbar \frac{dk_x}{dt} = g\mu_B\partial_xB -\frac{\epsilon_{k_x}}{T}\partial_x T,
\end{equation}
where $\epsilon_{k_x}$ is the dispersion of the chiral bands. For $\epsilon_0\simeq \mathcal K$, the velocities of the right-moving and left-moving chiral bands $v_{R,L}(\epsilon_{k_x})$ barely depend on $k_x$. Consequently, their magnon population edges are respectively shifted to $\epsilon_R=\epsilon_0+\epsilon_5$ and $\epsilon_L=\epsilon_0-\epsilon_5$ by the driving forces [Figs.~\ref{fig4}(c) and~\ref{fig4}(d)], with $\epsilon_5=\int \hbar v_Rdk_x = -\int \hbar v_Ldk_x$ resembling the chiral chemical potential in Weyl semimetals~\cite{zhang2016, burkov2015, liqiang2016, fukushima2008}. The shift of magnon population edges causes in the two chiral bands magnon concentration variations
\begin{align} \label{n}
n_{R,L} =\int_{\epsilon_{R,L}}^{\epsilon_0} \rho(\epsilon) n_B(\epsilon) d\epsilon \approx \frac{n_B(\epsilon_0)}{h v_R(\epsilon_0)}(\epsilon_0-\epsilon_{R,L}),
\end{align}
where $n_B(\epsilon)=(e^{\epsilon/k_BT}-1)^{-1}$ is the Bose-Einstein distribution function, $\rho(\epsilon)=1/hv_R(\epsilon)$ is the density of states, and the approximation stands in the limit $|\epsilon_5|\ll k_BT$. Labelling the  chiral charge of the left-moving band as $\chi$, the chirality imbalance reads $\rho_5=\chi(n_L-n_R)$, giving rise to 
\begin{equation} \label{mca}
\begin{split}
\frac{d\rho_5}{dt} \approx \chi \frac{n_B(\epsilon_0)}{\pi\hbar} \left(g\mu_B\partial_xB-\frac{\epsilon_{k_x}}{T}\partial_x T \right),
\end{split}
\end{equation}
which characterizes the (1+1)-dimensional magnon chiral anomaly. Unlike its (3+1)-dimensional counterpart~\cite{liu2019, su2017a, su2017b}, the (1+1)-dimensional magnon chiral anomaly does not involve the electric field gradient. In fact, the electric field gradient for magnons is dual to the magnetic field for electrons, which is absent in the (1+1)-dimensional spacetime~\cite{jackson1999}.

We now briefly discuss the magnon transport associated with the (1+1)-dimensional chiral anomaly. In the Boltzmann transport theory~\cite{suppl}, the magnon dipole conductivity and thermal conductivity respectively read
\begin{equation} \label{transport}
\begin{split}
\sigma(\epsilon_0)&=[\rho_M(\epsilon_0)]^{-1}= \frac{g^2\mu_B^2 \tau}{\pi\hbar}  v_R(\epsilon_0) n_B(\epsilon_0), 
\\
\kappa(\epsilon_0)&=[\rho_Q(\epsilon_0)]^{-1}= \frac{k_B^2T \tau}{\pi\hbar} v_R(\epsilon_0) n_B(\epsilon_0), 
\end{split}
\end{equation}
where $\tau$ is the relaxation time. Comparing to the electron Lorenz number $L_E=\pi^2k_B^2/3e^2$, a magnon Lorenz number can be defined as $L_M=\tfrac{\kappa}{T\sigma}=k_B^2/(g\mu_B)^2$~\cite{nakata2015, nakata2017}, where the magnon dipole moment $g\mu_B$ plays the role of the electron charge $-e$. In a strain-free ribbon, the magnon drift velocity $v_R(\epsilon_0)\simeq \sqrt{3}JS\delta_x/2\hbar$ is approximately constant, and both the dipole conductivity and the thermal conductivity grow rapidly with temperature [Fig.~\ref{fig4}(e)]. On the other hand, the magnon drift velocity of the strained ribbon reads $v_R(\epsilon_0)= 2^{\frac{5}{4}}3^{-\frac{7}{16}}(\epsilon_0-\mathcal K)^{\frac{1}{4}}  [(JS\sqrt{|\nu|\gamma\lambda\delta_x})^{\frac{3}{2}}- (\epsilon_0-\mathcal K)^{\frac{3}{2}} /(3^{\frac{9}{8}}\sqrt 2)]^{\frac{1}{2}} \delta_x/\hbar$, where $\epsilon_0$ intersects the $\nu$th magnon pseudo-Landau level. Since $v_R(\epsilon_0)$ proves to be an increasing function of the strain $\lambda$, the magnon dipole resistivity and thermal resistivity decrease with $\lambda$ [Fig.~\ref{fig4}(f)], leading to a negative strain-resistivity analogous to the negative magnetoresistivity~\cite{huang2015, kim2013, xiong2015, son2013, zhang2016, burkov2015, liqiang2016, fukushima2008} in the (3+1)-dimensional chiral anomaly of Weyl semimetals. Such a negative strain-resistivity may work as a direct manifestation of the (1+1)-dimensional magnon chiral anomaly [Eq.~(\ref{mca})]. 

\textit{Experimental accessibility.---} The implementation of the aforementioned (1+1)-dimensional magnon chiral anomaly first requires zigzag ribbons of two-dimensional honeycomb ferromagnets, which should be accessible in several van der Waals ferromagnets such as CrX$_3$ (X=Cl, I, Br)~\cite{chenlebing2022, caizhengwei2021, chenlebing2018, huang2017, chenlebing2021, kim2019, pershoguba2018} and CrXTe$_3$ (X=Si, Ge)~\cite{zhufengfeng2021}. In particular, both atomically thin sheets and ribbons of CrI$_3$ can be prepared through mechanical exfoliation~\cite{huang2017}, demonstrating the promise for the (1+1)-dimensional magnon chiral anomaly arising from the chiral bulk-edge magnon bands. Moreover, first-principles calculations have suggested superior flexibility in CrI$_3$, which may sustain tensile strain up to 10\%~\cite{wuzewen2019}. This makes CrI$_3$ a potentially ideal venue for the (1+1)-dimensional magnon chiral anomaly resulting from chiral magnon pseudo-Landau levels. Inelastic neutron scattering has been applied to study the magnonic band structures of CrI$_3$ \cite{chenlebing2018, chenlebing2021} as well as other honeycomb ferromagnets \cite{chenlebing2022, caizhengwei2021, zhufengfeng2021} and should be able to resolve both types of chiral magnon bands. The insulating nature of CrI$_3$ indicates that its transport should be mostly contributed by magnons and phonons \cite{seyler2018}. In spin caloritronic experiments, the magnon thermal conductivity could be extracted with the help of an applied magnetic field \cite{boona2014}. We thus expect the proposed (1+1)-dimensional magnon chiral anomaly to be tested readily by experiments.

The (3+1)-dimensional magnon chiral anomaly has been proposed in magnonic analogs of Weyl semimetals~\cite{mook2016, lifeiye2016, jian2018, lifeiye2017, owerre2018}, which may be realized by multilayer~\cite{su2017a} and pyrochlore~\cite{su2017b} ferromagnets. However, such magnon chiral anomaly in general requires an extremely inhomogeneous electric field (e.g., $|\bm \nabla \cdot \bm E|\sim 10^{22}\,$\text{V/m}$^2$ in Ref.~\cite{su2017a}) to induce chiral magnon Landau levels. In contrast, our chiral bulk-edge magnon bands and strain-induced chiral pseudo-Landau levels do not require nonuniform electric fields, thus in principle paving an easier way to the magnon chiral anomaly.

\textit{Conclusions.---} We show that edge and strain effects of two-dimensional honeycomb ferromagnets can push the chiral anomaly beyond the fermionic paradigm. The resulting (1+1)-dimensional magnon chiral anomaly is produced by either the chiral bulk-edge magnon bands arising from the edge-induced onsite energy modulation or the chiral magnon pseudo-Landau levels induced by strain. In particular, the (1+1)-dimensional magnon chiral anomaly supported by the pseudo-Landau levels features a negative strain-resistivity for both magnetic dipole and heat. Similar chiral anomalies may also be realized by Weyl photons~\cite{gaowenlong2016, guoqinghua2017} and Weyl phonons~\cite{xiaomeng2015, yangzhaoju2016, lifeng2018}, which may further enrich the quantum anomaly family.

\vspace{1em}
\begin{acknowledgments}
The authors are indebted to P. A. McClarty, R. Moessner, J.-H. Zhu, A. M. Cook, M. Vojta, H. Kondo, J. Sun, H.-M. Guo, S.-Q. Shen and J.-L. Chen for insightful discussions. This work was supported by the National Key R$\&$D Program of China (2022YFA1403700), Guangdong Basic and Applied Basic Research Foundation (2022A1515111034), the Max-Planck-Gesellschaft scholarship, the Innovation Program for Quantum Science and Technology (2021ZD0302400), the National Natural Science Foundation of China (11925402), Guangdong province (2020KCXTD001, 2016ZT06D348), the Science, Technology and Innovation Commission of Shenzhen Municipality (ZDSYS20170303165926217, JAY20170412152620376, and KYTDPT20181011104202253). The numerical calculations were supported by Center for Computational Science and Engineering of SUSTech.
\end{acknowledgments}

\bibliographystyle{apsrev4-1-etal-title_10authors}
\bibliography{MCA20230530}
\end{document}

\title{Supplemental Material for \\``Chiral Anomaly Beyond Fermionic Paradigm''}

\author{Tianyu Liu}
%\email{tliu@pks.mpg.de}
\affiliation{Shenzhen Institute for Quantum Science and Engineering and Department of Physics, Southern University of Science and Technology (SUSTech), Shenzhen 518055, China}
\affiliation{International Quantum Academy, Shenzhen 518048, China}
\affiliation{Max-Planck-Institut f\"ur Physik komplexer Systeme, 01187 Dresden, Germany}

\author{Zheng Shi}
%\email{zheng.shi@uwaterloo.ca}
\affiliation{Institute for Quantum Computing and Department of Physics and Astronomy, University of Waterloo, Waterloo, Ontario, N2L 3G1, Canada}
\affiliation{Dahlem Center for Complex Quantum Systems and Physics Department, Freie Universit\"at Berlin, 14195 Berlin, Germany}

\author{Hai-Zhou Lu}
\email{Corresponding author: luhz@sustech.edu.cn}
%\email{luhz@sustech.edu.cn}
\affiliation{Shenzhen Institute for Quantum Science and Engineering and Department of Physics, Southern University of Science and Technology (SUSTech), Shenzhen 518055, China}
\affiliation{Quantum Science Center of Guangdong-Hong Kong-Macao Greater Bay Area (Guangdong), Shenzhen 518045, China}
\affiliation{Shenzhen Key Laboratory of Quantum Science and Engineering, Shenzhen 518055, China}
\affiliation{International Quantum Academy, Shenzhen 518048, China}

\author{X. C. Xie}
\affiliation{International Center for Quantum Materials, School of Physics, Peking University, Beijing100871, China}
\affiliation{Institute for Nanoelectronic Devices and Quantum Computing, Fudan University, Shanghai 200433, China}
\affiliation{Hefei National Laboratory, Hefei 230088, China}

\date{\today}
\maketitle
\tableofcontents

\section{Periodic model with next-nearest-neighbor interactions}
\label{s1}
In the main text, we characterize the honeycomb ferromagnets with a  simple Heisenberg Hamiltonian [Eq.~(\textcolor{blue}{1}) of the main text] containing only nearest-neighbor terms and onsite terms. We here consider a more generic model with also the second-nearest-neighbor terms. Explicitly, the model reads
\begin{equation} \label{generic}
H= - \sum_{\langle i,j\rangle} J_{ij}\bm S_i \cdot \bm S_j - \sum_i K_i(\bm S_i\cdot \hat z)^2 - \sum_{\langle\langle i,j \rangle\rangle} J'_{ij}\bm S_i \cdot \bm S_j + \sum_{\langle\langle i,j\rangle\rangle} \bm D_{ij}\cdot (\bm S_i \times \bm S_j), 
\end{equation}
where the first two terms are the same as Eq.~(\textcolor{blue}{1}) of the main text, the third term represents the next-nearest-neighbor exchange interaction with $J'_{ij}>0$ because of the ferromagnetism, and the last term is the next-nearest-neighbor Dyzaloshinskii-Moriya interaction \cite{dzyaloshinskii1958, moriya1960} with a vector strength $\bm D_{ij}$. 

\begin{figure} [tb]
\includegraphics[width = 8.6cm]{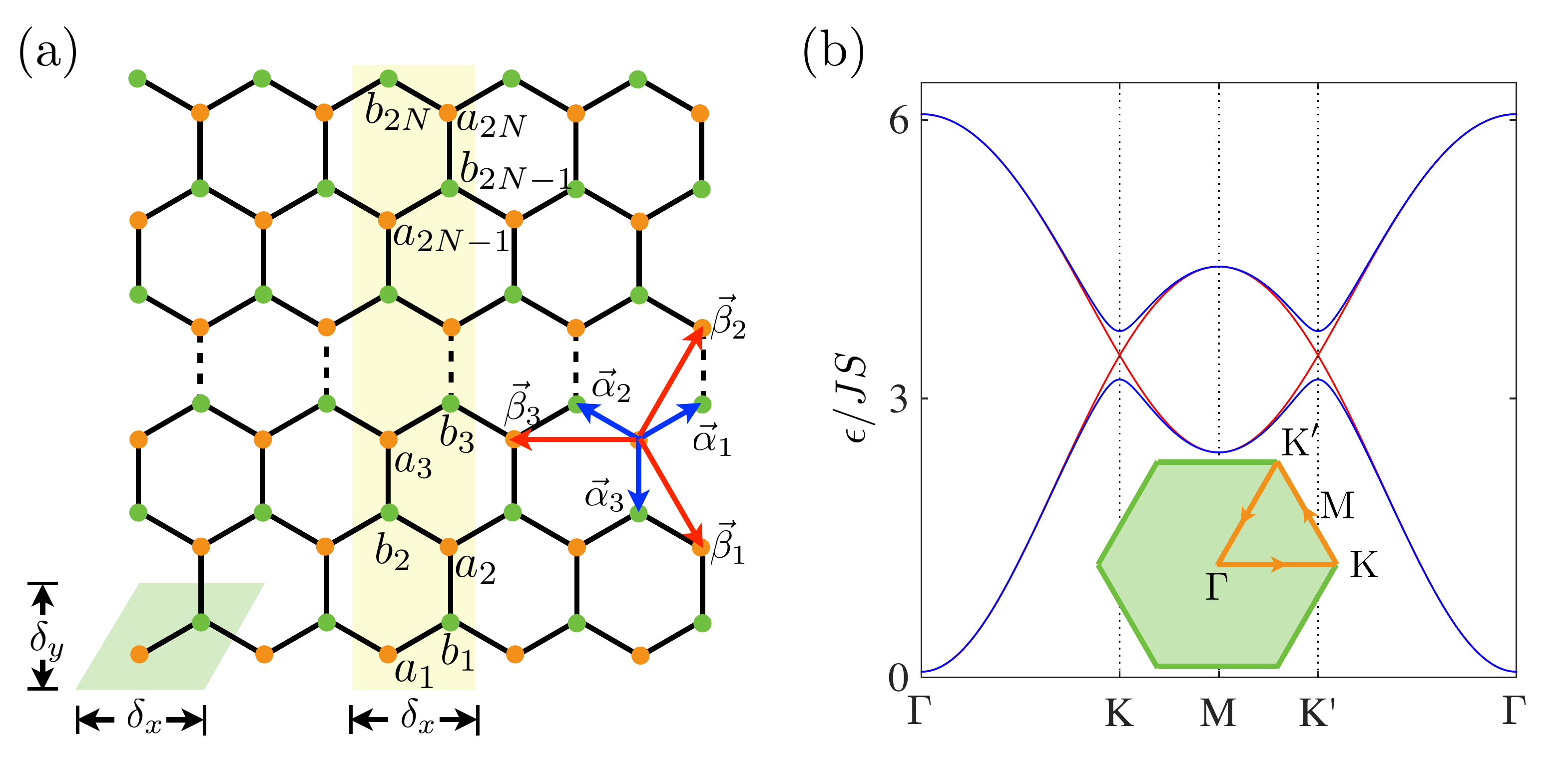}
\caption{(a) A honeycomb lattice of spins. The nearest-neighbor vectors $\bm \alpha_{1,2,3}$ and next-nearest-neighbor vectors $\bm \beta_{1,2,3}$ are labeled by blue and red arrows, respectively. For the periodic model, the unit cell contains an $A$-sublattice site and a $B$-sublattice site and is marked by the green rhombic shade with width $\delta_x$ and height $\delta_y$. For the ribbon model, we adopt a unit cell (yellow shade of width $\delta_x$) of $2N$ $A$-sublattice sites and the same number of $B$-sublattice sites. (b) Magnon spectrum [Eq.~(\ref{spec_periodic})] of the periodic model of the honeycomb ferromagnet plotted along a path connecting the high-symmetry points (orange). The spectrum without the Dzyaloshinskii-Moriya interaction (red) exhibits a pair of Dirac points at the corners of the hexagonal Brillouin zone (green), while a finite Dzyaloshinskii-Moriya interaction gaps out the Dirac cones (blue). The parameters are set by referring to CrBr$_3$ \cite{caizhengwei2021} as $K=0.03J$, $J'=0.045J$, $D=0.05J$. These parameters are also comparable to those of other honeycomb ferromagnets \cite{chenlebing2018, zhufengfeng2021}
} \label{figs1}
\end{figure}

We now analyze the magnonic structure of the generic model [Eq.~(\ref{generic})] defined on a spin-$S$ honeycomb lattice [Fig.~\ref{figs1}(a)]. For this purpose, we rewrite Eq.~(\ref{generic}) with the nearest-neighbor and next-nearest-neighbor vectors, which respectively read $(\bm \alpha_1, \bm \alpha_2, \bm \alpha_3) = (\tfrac{\sqrt 3}{2}a \hat x + \tfrac{1}{2}a \hat y, -\tfrac{\sqrt 3}{2}a \hat x + \tfrac{1}{2}a \hat y, -a\hat y)$ and $(\bm \beta_1, \bm \beta_2, \bm \beta_3) = (\tfrac{\sqrt 3}{2}a \hat x - \tfrac{3}{2}a \hat y, \tfrac{\sqrt 3}{2}a \hat x + \tfrac{3}{2}a \hat y, -\sqrt 3 a\hat x)$ with $a$ being the lattice constant. After rewriting, the model Hamiltonian becomes
\begin{equation} \label{H_periodic}
H = -\sum_{\bm r,j} J_j(\bm r) \bm S_a(\bm r) \cdot \bm S_b(\bm r + \bm \alpha_j-\bm \alpha_1) - \sum_{\bm r,\mu} K(\bm r)[S_\mu^z(\bm r)]^2 + \sum_{\bm r, j,\mu} \bm D_\mu^j(\bm r) \cdot [\bm S_\mu(\bm r) \times \bm S_\mu(\bm r + \bm \beta_j)]  -\sum_{\bm r,j,\mu}  J'_j(\bm r) \bm S_\mu(\bm r) \cdot \bm S_\mu (\bm r + \bm \beta_j), 
\end{equation}
where $\mu=a, b$ is the sublattice index, $j=1,2,3$ specifies the nearest-neighbor and next-nearest-neighbor vectors, and $\bm r=(x,y)$ labels the position of a unit cell comprising an $A$-sublattice site and a $B$-sublattice site connected by $\bm \alpha_1$ [green shade, Fig.~\ref{figs1}(a)]. For the candidate honeycomb ferromagnets (e.g., CrBr$_3$ \cite{caizhengwei2021}, CrI$_3$ \cite{chenlebing2018} and CrXTe$_3$ (X=Si, Ge) \cite{zhufengfeng2021}) in consideration, it would be sufficient to choose $K(\bm r)=K$, $J_j(\bm r)=J$, $J'_j(\bm r)=J'$ and $\bm D_b^j(\bm r)=-\bm D_a^j(\bm r)=D\hat z$.

The second quantization of the Heisenberg Hamiltonian [Eq.~(\ref{H_periodic})] can be performed by applying the Holstein-Primakoff transformation \cite{holstein1940} $S_\mu^+(\bm r) = [S_\mu^-(\bm r)]^\dagger = (2S-\mu_{\bm r}^\dagger \mu_{\bm r})^{1/2} \mu_{\bm r}$ and $S_\mu^z(\bm r) = S-\mu_{\bm r}^\dagger \mu_{\bm r}$, where $\mu_{\bm r}^\dagger$/$\mu_{\bm r}$ is the magnon creation/annihilation operator on sublattice $\mu=a,b$. In the large-$S$ limit, the magnon-magnon interactions can be ignored, leading to a linear spin wave theory
\begin{equation} \label{tb_periodic}
H = \mathcal K \sum_{\bm r} (a_{\bm r}^\dagger a_{\bm r} +b_{\bm r}^\dagger b_{\bm r}) + \sum_{\bm r,j} \left[  -JS a_{\bm r}^\dagger b_{\bm r+\bm \alpha_j-\bm\alpha_1} -J'S (a_{\bm r}^\dagger a_{\bm r+\bm \beta_j} + b_{\bm r}^\dagger b_{\bm r+\bm \beta_j} ) + iDS (a_{\bm r}^\dagger a_{\bm r+\bm \beta_j}-b_{\bm r}^\dagger b_{\bm r+\bm \beta_j}) + \text{H.c.} \right],
\end{equation}
where we define $\mathcal K=2KS+3JS+6J'S$, subtract from Eq.~(\ref{H_periodic}) the ferromagnetic ground state energy $E_G=-N_{\text{uc}}\mathcal KS$ with $N_{\text{uc}}$ being the number of unit cells, and make use of $\sum_{\bm r}\mu_{\bm r+\bm \beta_i}^\dagger \mu_{\bm r+\bm \beta_i}=\sum_{\bm r}\mu_{\bm r}^\dagger \mu_{\bm r}$  under periodic boundary conditions. Eqnation~(\ref{tb_periodic}) is a magnon tight-binding Hamiltonian, where $-J$ ($-J'$) is the real nearest-neighbor (next-nearest-neighbor) hopping, and the imaginary next-nearest-neighbor hopping $-iDS$ ($+iDS$) along $\bm \beta_j$ on the $A$ ($B$) sublattice is analogous to the intrinsic spin-orbit coupling in graphene \cite{kane2005}. Besides the easy-axis anisotropy, we note that the onsite energy is also associated with the hoppings $-J$ and $-J'$ and depends on the numbers of nearest neighbors and next-nearest neighbors of a lattice site. Once we perform the Fourier transform, $(a_{\bm r}, b_{\bm r})^T=N_{\text{uc}}^{-1/2}\sum_{\bm k} e^{i\bm k\cdot \bm r} (a_{\bm k}, e^{i\bm k\cdot \bm \alpha_1}b_{\bm k})^T$, the magnon tight-binding Hamiltonian can be written in the sublattice basis $\psi_{\bm k} = (a_{\bm k}, b_{\bm k})^T$ as $H=\sum_{\bm k} \psi_{\bm k}^\dagger \mathcal H_{\bm k} \psi_{\bm k}$ with the following Bloch Hamiltonian
\begin{equation} \label{Bloch_periodic}
\mathcal H_{\bm k} = \left[ \mathcal K-2J'S\sum_j \cos(\bm k \cdot \bm \beta_j)\right] \sigma^0 -JS \sum_j \cos(\bm k \cdot \bm \alpha_j) \sigma^x + JS \sum_j \sin(\bm k \cdot \bm \alpha_j) \sigma^y -2DS \sum_j \sin(\bm k \cdot \bm \beta_j) \sigma^z,
\end{equation}
where $\sigma^{x,y,z}$ and $\sigma^0$ are the Pauli matrices and the identity matrix in the sublattice space, respectively. This magnon Bloch Hamiltonian can be unitarily diagonalized to find the spin wave dispersion as
\begin{equation} \label{spec_periodic}
\epsilon_{\bm k} = 2KS+3JS+6J'S-2J' S \sum_j \cos(\bm k \cdot \bm \beta_j) \pm d_{\bm k},
\end{equation}
with
\begin{equation} \label{d_periodic}
d_{\bm k} =  \sqrt{J^2S^2\left[1+4\cos^2\left(\frac{1}{2}k_x\delta_x\right)+4\cos\left(\frac{1}{2}k_x\delta_x\right) \cos(k_y\delta_y)\right] +4D^2S^2 \left[\sin(k_x\delta_x)-2\sin\left(\frac{1}{2}k_x\delta_x\right)\cos(k_y\delta_y)\right]^2},
\end{equation}
where $\delta_x=\sqrt 3a$ and $\delta_y=\tfrac{3}{2}a$ measure the size of the unit cell of the periodic model [Fig.~\ref{figs1}(a)]. In the absence of the Dzyaloshinskii-Moriya interaction, Eq.~(\ref{Bloch_periodic}) is reduced to a standard Dirac Hamiltonian exhibiting a pair of Dirac cones at the Brillouin zone corners $K$ and $K'$ [Fig.~\ref{figs1}(b)]. The Dzyaloshinskii-Moriya interaction opens up gaps $\Delta_K=-6\sqrt 3 DS$ and $\Delta_{K'}=6\sqrt 3 DS$ at $K$ and $K'$, respectively. This band gap inversion indicates the nontrivial topology embedded in $\mathcal H_{\bm k}$, which can be regarded as a magnonic implementation of the Haldane model \cite{haldane1988}, characterized by a Chern number $\mathcal C=\pm\tfrac{1}{2}[\text{sgn}(\Delta_{K'})-\text{sgn}(\Delta_K)]=\pm 1$. The bulk-boundary correspondence consequently suggests a pair of in-gap chiral magnon states on the edges of the honeycomb ferromagnet.

\section{Ribbon model with next-nearest-neighbor interactions}
\label{s2}
In Sec.~\ref{s1}, we have studied a generic periodic model with next-nearest-neighbor interactions [Eq.~(\ref{H_periodic})]. We here derive the form of the generic model for a zigzag ribbon to justify the Bloch Hamiltonian Eq.~(\textcolor{blue}{3}) of the main text.

We choose the unit cell of the ribbon to contain $2N$ $A$-sublattice sites and the same number of $B$-sublattice sites [yellow shade, Fig.~\ref{figs1}(a)]. The resulting one-dimensional Heisenberg Hamiltonian is
\begin{equation} \label{H_ribbon}
H=\sum_x (H_0+H_1+H_2+H_D),
\end{equation}
with its components listed explicitly below:
\begin{align}
H_0 = &-K\sum_{j=1}^{2N}\left\{[\hat z\cdot \bm S_a^j(x)]^2+[\hat z \cdot \bm S_b^j(x)]^2\right\},
\\
H_1= &-J \sum_{j=1}^{2N} \bm S_a^j (x) \cdot \bm S_b^j (x)-J \sum_{j=1}^{2N-1} \bm S_b^j (x) \cdot  \bm S_a^{j+1} (x) -J \sum_{i=1}^{N} \bm S_b^{2i-1} (x) \cdot  \bm S_a^{2i-1} (x+\delta_x) \nonumber
\\
&-J \sum_{i=1}^{N} \bm S_a^{2i} (x) \cdot  \bm S_b^{2i} (x+\delta_x), 
\\
H_2 = & -J' \sum_{j=1}^{2N} \left[ \bm S_a^j(x) \cdot \bm S_a^j(x-\delta_x) + \bm S_b^j(x) \cdot \bm S_b^j(x-\delta_x) \right] - J' \sum_{i=1}^{N-1} \bm S_a^{2i}(x) \cdot \left[\bm S_a^{2i+1}(x)  +  \bm S_a^{2i+1}(x+\delta_x)\right] \nonumber
\\
&- J'  \sum_{i=1}^{N} \bm S_a^{2i}(x) \cdot \left[ \bm S_a^{2i-1}(x) +  \bm S_a^{2i-1}(x+\delta_x)\right] -J'  \sum_{i=1}^{N-1} \bm S_b^{2i+1}(x) \cdot \left[\bm S_b^{2i}(x+\delta_x) + \bm S_b^{2i}(x)\right] \nonumber
\\
&-J'  \sum_{i=1}^{N} \bm S_b^{2i-1}(x) \cdot \left[\bm S_b^{2i}(x+\delta_x) + \bm S_b^{2i}(x)\right],
\\
H_D =& D \sum_{j=1}^{2N} \hat z \cdot \left[\bm S_a^j(x-\delta_x)\times \bm S_a^j(x) - \bm S_b^j(x-\delta_x)\times \bm S_b^j(x) \right] + D \sum_{i=1}^{N-1} \hat z \cdot \left\{ \bm S_a^{2i}(x) \times \left[\bm S_a^{2i+1}(x) -\bm S_a^{2i+1}(x+\delta_x)\right] \right\} \nonumber
\\
+& D \sum_{i=1}^{N} \hat z \cdot \left\{ \bm S_a^{2i}(x)\times \left[\bm S_a^{2i-1}(x) - \bm S_a^{2i-1}(x+\delta_x)\right] \right\} + D \sum_{i=1}^{N-1} \hat z  \cdot \left\{ \bm S_b^{2i+1}(x) \times \left[\bm S_b^{2i}(x+\delta_x)- \bm S_b^{2i}(x) \right] \right\} \nonumber
\\
+& D \sum_{i=1}^{N} \hat z \cdot  \left\{ \bm S_b^{2i-1}(x) \times \left[\bm S_b^{2i}(x+\delta_x)- \bm S_b^{2i}(x)\right] \right\},
\end{align}
where $H_0$ is the easy-axis anisotropy, $H_1$ is the nearest-neighbor exchange interaction, $H_2$ is the next-nearest-neighbor exchange interaction, and $H_D$ is the next-nearest-neighbor Dzyaloshinskii-Moriya interaction.

The second quantization of the ribbon Hamiltonian [Eq.~(\ref{H_ribbon})] can be implemented by the Holstein-Primakoff transformation $S_\mu^{j,+}(x) = [S_\mu^{j,-}(x)]^\dagger= \sqrt{2S}\mu_{x,j}$, $S_\mu^{j,z}(x) = S-\mu_{x,j}^\dagger\mu_{x,j}$ in the large-$S$ limit with $\mu_{x,j}^\dagger/\mu_{x,j}$ being the magnon creation/annihilation operator on the sublattice $\mu=a,b$. With the ferromagnetic ground state energy subtracted, the components of the ribbon Hamiltonian become
\begin{align}
H_0=2KS &\sum_{j=1}^{2N}\left(a_{x,j}^\dagger a_{x,j}+b_{x,j}^\dagger b_{x,j} \right),
\\
H_1= 3JS &\sum_{j=2}^{2N-1} (a_{x,j}^\dagger a_{x,j}+b_{x,j}^\dagger b_{x,j}) +2JS (a_{x,1}^\dagger a_{x,1} + b_{x,2N}^\dagger b_{x,2N}) + 3JS (a_{x,2N}^\dagger a_{x,2N} + b_{x,1}^\dagger b_{x,1}) \nonumber
\\
-JS &\left[\sum_{i=1}^{N} (b_{x,2i-1}^\dagger a_{x+\delta_x,2i-1} + a_{x,2i}^\dagger b_{x+\delta_x,2i}) +\sum_{j=1}^{2N} a_{x,j}^\dagger b_{x,j} + \sum_{j=1}^{2N-1} a_{x,j+1}^\dagger b_{x,j} + \text{H.c.} \right],
\\
H_2= 6J'S &\sum_{j=2}^{2N-1} (a_{x,j}^\dagger a_{x,j}+b_{x,j}^\dagger b_{x,j}) + 4J'S (a_{x,1}^\dagger a_{x,1} + b_{x,1}^\dagger b_{x,1} + a_{x,2N}^\dagger a_{x,2N} + b_{x,2N}^\dagger b_{x,2N}) \nonumber
\\
-J'S &\left[ \sum_{j=1}^{2N} \left(a_{x,j}^\dagger a_{x-\delta_x,j} + b_{x,j}^\dagger b_{x-\delta_x,j} \right) + \sum_{i=1}^{N-1} a_{x,2i}^\dagger \left(a_{x, 2i+1}  +  a_{x+\delta_x,2i+1}\right) + \sum_{i=1}^N  a_{x,2i}^\dagger \left(a_{x, 2i-1}  + a_{x+\delta_x,2i-1} \right) \right. \nonumber
\\
+ &\left.\sum_{i=1}^{N-1} b_{x,2i+1}^\dagger \left(b_{x,2i}  + b_{x+\delta_x,2i}\right) + \sum_{i=1}^N b_{x,2i-1}^\dagger \left(b_{x,2i} + b_{x+\delta_x,2i}\right) +\text{H.c.} \right],
\\
H_D = iDS &\left[ \sum_{j=1}^{2N} \left( a_{x,j}^\dagger a_{x-\delta_x,j}-b_{x,j}^\dagger b_{x-\delta_x,j} \right) + \sum_{i=1}^{N-1} a_{x,2i}^\dagger \left(a_{x+\delta_x,2i+1} -  a_{x,2i+1} \right) + \sum_{i=1}^N a_{x,2i}^\dagger \left(a_{x+\delta_x,2i-1}
-  a_{x,2i-1} \right)\right. \nonumber
\\
+ &\left.\sum_{i=1}^{N-1} b_{x,2i+1}^\dagger \left( b_{x,2i}  - b_{x+\delta_x,2i} \right) + \sum_{i=1}^N b_{x,2i-1}^\dagger \left(b_{x,2i} -  b_{x+\delta_x,2i}\right)+ \text{H.c.} \right],
\end{align}
where the nearest-neighbor hopping is $-JS$ and the next-nearest-neighbor hopping associated with the $A$ ($B$) sublattice is $-J'S \mp iDS$ ($-J'S \pm iDS$) along $\pm \bm \beta_i$. Along the edge direction, the translational symmetry is preserved, allowing the partial Fourier transform
\begin{equation} \label{nanoribbon_FT}
\begin{pmatrix}
a_{x, 2i-1} \\ a_{x, 2i} \\ b_{x, 2i-1} \\ b_{x, 2i} 
\end{pmatrix}
= \frac{1}{\sqrt {\tilde N_{\text{uc}}} }\sum_{k_x} e^{ik_xx} 
\begin{pmatrix}
e^{-ik_x\delta_x/2} a_{k_x, 2i-1}
\\
a_{k_x, 2i}
\\
b_{k_x, 2i-1}
\\
e^{-ik_x\delta_x/2} b_{k_x, 2i}
\end{pmatrix},
\end{equation}
where $\tilde N_{\text{uc}}$ is the number of the ribbon unit cells [yellow shade, Fig.~\ref{figs1}(a)]. Afterwards, the magnon tight-binding Hamiltonian of the zigzag ribbon can be arranged in a compact form $H=\sum_{k_x} \Psi_{k_x}^\dagger \mathcal H_{k_x} \Psi_{k_x}$ with the basis $\Psi_{k_x} = (a_{k_x,1}, \cdots, a_{k_x,2N}, b_{k_x,1}, \cdots, b_{k_x,2N})^T$ and the ribbon Bloch Hamiltonian
\begin{equation}  \label{Bloch_ribbon}
\mathcal H_{k_x} = 
\renewcommand*{\arraystretch}{1.5}
\begin{pmatrix}
\mathcal H_{k_x}^{AA} & \mathcal H_{k_x}^{AB}
\\
\mathcal H_{k_x}^{BA} & \mathcal H_{k_x}^{BB}
\end{pmatrix},
\end{equation}
which comprises of four $2N\times 2N$ blocks. The diagonal blocks $\mathcal H_{k_x}^{AA}$ and $\mathcal H_{k_x}^{BB}$ contain the next-nearest-neighbor terms arising from the exchange interaction and the Dzyaloshinskii-Moriya interaction as well as the onsite terms resulting from the nearest-neighbor and next-nearest-neighbor exchange interactions and easy-axis anisotropy. The off-diagonal blocks $\mathcal H_{k_x}^{AB}$ and $\mathcal H_{k_x}^{BA}$ enclose the information of the nearest-neighbor exchange interaction. Explicitly, these terms are 
\begin{subequations} \label{H_comp}
\begin{align}
\mathcal H_{k_x}^{AA} = &-\left[2J'S \cos\left(\frac{1}{2}k_x\delta_x\right) + 2DS \sin\left(\frac{1}{2}k_x\delta_x\right)\right] (\bm L + \bm L^\dagger)  - [2J'S \cos(k_x\delta_x) - 2DS \sin(k_x\delta_x)] \bm I \nonumber
\\
&+2KS\bm I+JS\bm \Lambda_A + J'S\bm \Lambda', 
\\
\mathcal H_{k_x}^{BB}= & - \left[2J'S \cos\left(\frac{1}{2}k_x\delta_x\right) - 2DS \sin\left(\frac{1}{2}k_x\delta_x\right)\right] (\bm L + \bm L^\dagger)  - [2J' S\cos(k_x\delta_x) + 2DS \sin(k_x\delta_x) ]\bm I \nonumber
\\
& +2KS\bm I+ JS\bm \Lambda_B + J'S \bm \Lambda',
\\
\mathcal H_{k_x}^{AB} = & -2JS\cos\left(\frac{1}{2}k_x\delta_x\right) \bm I - JS \bm L,
\\
\mathcal H_{k_x}^{BA} = &-2JS\cos\left(\frac{1}{2}k_x\delta_x\right) \bm I - JS \bm L^\dagger,
\end{align}
\end{subequations}
where $\bm I$ is the identity matrix and we have defined for transparency the following $2N\times2N$ matrices 
\begin{align} \label{mat}
\bm L = 
\begin{pmatrix}
0 & \cdots & 0 &0 & 0
\\
1 & \cdots & 0 & 0 & 0
\\
\vdots & \ddots & \vdots & \vdots & \vdots
\\
0 & \cdots &  1& 0 & 0
\\
0 & \cdots &  0& 1 & 0
\end{pmatrix},
\quad 
\bm \Lambda_A = 
\begin{pmatrix}
2 & 0 & \cdots & 0 & 0
\\
0 & 3 & \cdots & 0 &0
\\
\vdots & \vdots & \ddots & \vdots & \vdots
\\
0 & 0 & \cdots & 3 & 0
\\
0 & 0 & \cdots & 0 & 3
\end{pmatrix}, 
\quad
\bm \Lambda' = 
\begin{pmatrix}
4 & 0 & \cdots & 0 & 0 
\\
0 & 6 & \cdots & 0 & 0 
\\
\vdots & \vdots & \ddots & \vdots & \vdots
\\
0 & 0 & \cdots & 6 & 0 
\\
0 & 0 & \cdots & 0 & 4
\end{pmatrix},
\quad
\bm \Lambda_B = 
\begin{pmatrix}
3 & 0 & \cdots & 0 & 0
\\
0 & 3 & \cdots & 0 & 0
\\
\vdots & \vdots & \ddots & \vdots & \vdots
\\
0 & 0 & \cdots & 3 & 0
\\
0 & 0 & \cdots & 0 & 2
\end{pmatrix}.
\end{align}
In Eq.~(\ref{mat}), matrices $\bm \Lambda_A$, $\bm \Lambda_B$, and $\bm \Lambda'$ are diagonal and thus are associated with the onsite energies of magnons. The edge sites [i.e., $a_1$, $b_1$, $a_{2N}$, and $b_{2N}$ in Fig.~\ref{figs1}(a)] exhibit distinct onsite energies, because their numbers of nearest and next-nearest neighbors differ from those of the bulk sites. Specifically, the $a_1$ site and the $b_{2N}$ site each has two nearest neighbors, while all the other sites each has three nearest neighbors, justifying the entries in $\bm \Lambda_A$ and $\bm \Lambda_B$. Moreover, sites $a_1$, $b_1$, $a_{2N}$, and $b_{2N}$ each has four next-nearest neighbors, but each of the bulk site has six next-nearest neighbors, reflecting the entries of $\bm \Lambda'$. In the absence of next-nearest-neighbor interactions (i.e., $J'=D=0$), Eq.~(\ref{Bloch_ribbon}) is reduced to Eq.~(\textcolor{blue}3) of the main text.

\section{Chiral magnon bands in zigzag ribbons}
\label{s3}
In Sec.~\ref{s2}, we have explicitly derived the Bloch Hamiltonian for a zigzag ribbon of honeycomb ferromagnets. We now investigate the spin-wave structure of the ribbon from both numerical and analytical aspects.
 
\begin{figure*} [t]
\includegraphics[width = 17.2cm]{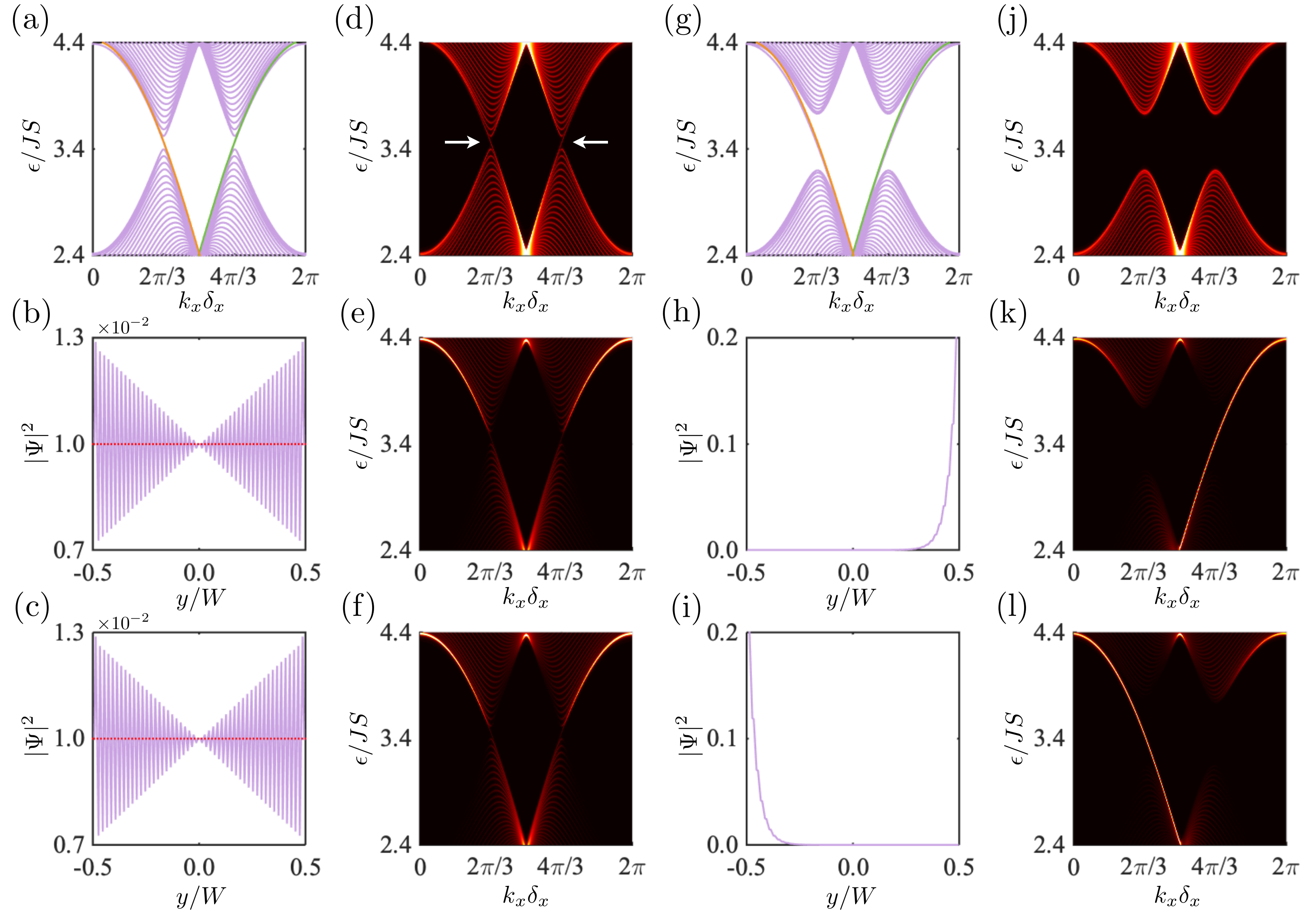}
\caption{Chiral magnon bands in the zigzag ribbon of honeycomb ferromagnets. The spectral properties in the absence and presence of the Dzyaloshinskii-Moriya interaction are plotted in (a)--(f) and (g)--(l), respectively. (a) Magnon dispersion of the zigzag ribbon (purple) and the overlaid analytic dispersions of the chiral magnon bands [Eq.~(\ref{spec_edge}), green and orange]. (b) Magnon probability distribution (purple) of the right-moving chiral magnon band at $k_x\delta_x=4\pi/3$. (c) Magnon probability distribution (purple) of the left-moving chiral magnon band at $k_x\delta_x=2\pi/3$. For panels (b) and (c), the red dotted lines represent the $J'=0$ distributions $|\Psi_R|^2$ [Eq.~(\ref{wf_A})] and $|\Psi_L|^2$ [Eq.~(\ref{wf_B})], respectively. (d) Spectral function of the bulk with the white arrows marking the chiral magnon bands. (e) Spectral function of the upper edge. (f) Spectral function of the lower edge. In panel (d), the $n$ sum in Eq.~(\ref{LDOS}) runs over 50 percent of lattice sites closest to the center of the ribbon, while (e) and (f) each includes 10 percent of lattice sites closest to the corresponding edge. (g) Magnon dispersion (purple) with a band gap opened by the Dzyaloshinskii-Moriya interaction. The analytic dispersions [Eq.~(\ref{spec_edge})] of the in-gap chiral magnon bands are overlaid as green and orange curves. (h) Magnon probability distribution of the right-moving chiral band at $k_x\delta_x=4\pi/3$. (i) Magnon probability distribution of the left-moving chiral band at $k_x\delta_x=2\pi/3$. (j) Spectral function of the bulk. (k) Spectral function of the upper edge. (l) Spectral function of the lower edge. We set $N=25$, $J'=0.045J$, $K=0.03J$ and, for panels (g)--(l) only, $D=0.05J$. 
} \label{figs2}
\end{figure*}

The exact diagonalization of the ribbon Bloch Hamiltonian [Eq.~(\ref{Bloch_ribbon})] reveals a pair of chiral magnon bands both in the absence [Fig.~\ref{figs2}(a)] and in the presence [Fig.~\ref{figs2}(g)] of the Dzyaloshinskii-Moriya interaction. To better understand the nature of these chiral magnon bands, we first calculate the spectral function through
\begin{equation} \label{LDOS}
A(k_x, \epsilon) = -\frac{1}{\pi}\sum_n \lim_{\delta \rightarrow 0} \Im [\epsilon + i\delta - \mathcal H_{k_x}^{nm}]^{-1}_{n=m}.
\end{equation}
We find that the chiral magnon bands in the absence of the Dzyaloshinskii-Moriya interaction are pronounced both in the bulk [Fig.~\ref{figs2}(d)] and on the edges [Figs.~\ref{figs2}(e) and~\ref{figs2}(f)]. In contrast, the chiral magnon bands are pushed out of the bulk by the Dzyaloshinskii-Moriya interaction [Fig.~\ref{figs2}(j)] and can only appear on the zigzag edges [Figs.~\ref{figs2}(k) and~\ref{figs2}(l)]. Such real-space features of the chiral magnon bands can be further confirmed by inspecting their wave functions. For a vanishing Dzyaloshinskii-Moriya interaction, the wave function of the right-moving (left-moving) chiral magnon band calculated at the momentum $k_x=\frac{4\pi}{3\delta_x}$ ($k_x=\frac{2\pi}{3\delta_x}$) depends strongly on the sublattice as illustrated in Fig.~\ref{figs2}(b) [Fig.~\ref{figs2}(c)], but remains finite across the whole ribbon along the $y$ direction, in agreement with the emergence of the right-moving (left-moving) chiral magnon bands in both the bulk spectral function [Fig.~\ref{figs2}(d)] and the edge spectral functions [Figs.~\ref{figs2}(e) and~\ref{figs2}(f)]. For this reason, we refer to the chiral magnon bands in the absence of Dzyaloshinskii-Moriya interaction as the bulk-edge states, which are generalizations of the chiral bulk-edge magnon bands discussed in the main text. In fact, the extra next-nearest-neighbor exchange interaction considered here only modifies the whole band structure with a shift and a tilt [see the third and fourth terms of Eq.~(\ref{spec_periodic})]. On the other hand, in the presence of the Dzyaloshinskii-Moriya interaction, we find the wave function of the right-moving (left-moving) chiral magnon band  calculated at the momentum $k_x=\frac{4\pi}{3\delta_x}$ ($k_x=\frac{2\pi}{3\delta_x}$) is localized around the upper (lower) zigzag edge of the ribbon, as illustrated in Fig.~\ref{figs2}(h) [Fig.~\ref{figs2}(i)], consistent with our observations in Figs.~\ref{figs2}(j), \ref{figs2}(k) and~\ref{figs2}(l). Such in-gap chiral magnon bands are unambiguously the topological edge states of the magnon Haldane model [Eq.~(\ref{Bloch_periodic})]. The chiral bulk-edge magnon bands and the chiral edge magnon bands are analogous to the chiral zeroth Landau levels of Weyl semimetals, and thus may result in similar transport features such as chiral anomaly. It is worth noting that the associated magnon chiral anomaly is defined in the (1+1)-dimensional spacetime spanned by the ribbon [cf.,  the (3+1)-dimensional spacetime spanned by Weyl semimetals].

To systematically study the transport, it is desirable to understand the dispersions of the chiral magnon bands in an analytic fashion. We first consider the chiral bulk-edge magnon bands emerging in the absence of the Dzyaloshinskii-Moriya interaction. We assume that the nearest-neighbor exchange interaction dominates the next-nearest-neighbor exchange interaction, i.e., $J\gg J'$, which is indeed the case in candidate honeycomb ferromagnets CrBr$_3$ \cite{caizhengwei2021}, CrI$_3$ \cite{chenlebing2018} and CrXTe$_3$ (X=Si, Ge) \cite{zhufengfeng2021}. Such an assumption allows us to determine the analytic dispersions of the bulk-edge bands through perturbation. Specifically, we first consider the eigenvalue problem of $\mathcal H_{k_x}^{(0)}=\mathcal H_{k_x}|_{D=J'=0}$, which is governed by the following Harper's equations
\begin{subequations}\label{harper}
\begin{align}
-JS\psi_{n-1}^B - 2JS\cos\left(\frac{1}{2}k_x\delta_x\right)\psi_n^B &= \left[\epsilon_{k_x}^{(0)} - 2KS - 3JS\right] \psi_n^A, \label{harper_A}
\\
-2JS\cos\left(\frac{1}{2}k_x\delta_x\right) \psi_m^A - JS\psi_{m+1}^A &= \left[\epsilon_{k_x}^{(0)} - 2KS - 3JS\right]  \psi_m^B, \label{harper_B}
\end{align}
\end{subequations}
where $(\psi_1^A, \cdots, \psi_{2N}^A, \psi_1^B, \cdots, \psi_{2N}^B)^T$ is the eigenvector of the matrix $\mathcal H_{k_x}^{(0)}$ corresponding to the eigenenergy $\epsilon_{k_x}^{(0)}$. Owing to the edge modulation reflected in matrices $\bm \Lambda_A$ and $\bm \Lambda_B$, in Eq.~(\ref{harper}) the indices $n=2,\cdots, 2N$ and $m=1,\cdots,2N-1$ exclude the edge sites $a_1$ and $b_{2N}$, whose onsite energy is $2KS+2JS$ rather than $2KS+3JS$. To solve the Harper's equations, we make an ansatz 
\begin{equation} \label{ansatz}
\begin{pmatrix}
\psi_{n}^A
\\
\psi_{n}^B
\end{pmatrix}
= z^n
\begin{pmatrix}
c_A
\\
c_B
\end{pmatrix},
\end{equation}
which represents a generalized Bloch theorem \cite{you2008}.  The existence of a nonzero solution to the Harper's equations [Eq.~(\ref{harper})] requires
\begin{equation} \label{d_ribbon}
\epsilon_{k_x}^{(0)}= 2KS+ 3JS \pm JS\sqrt{1 + 4\cos^2\left(\frac{1}{2}k_x\delta_x\right) + 2 \cos\left(\frac{1}{2}k_x\delta_x\right)(z+z^{-1})}.
\end{equation}
For the periodic model [Eq.~(\ref{Bloch_periodic})], the usual Bloch theorem is restored and the parameter $z$ characterizes a plane wave $z=e^{\pm ik_y\delta_y}$; Equation~(\ref{d_ribbon}) is then the same as Eq.~(\ref{spec_periodic}), provided that the next-nearest-neighbor interactions in Eq.~(\ref{spec_periodic}) are turned off, i.e., $D=J'=0$. Because of the edge modulation of the zigzag ribbon, $z$ in general adopts a different value from $e^{\pm ik_y\delta_y}$ and is governed by the boundary conditions 
\begin{subequations} \label{boundary}
\begin{align}
 - 2JS \cos\left(\frac{1}{2}k_x\delta_x\right) \psi_1^B &= \left[\epsilon_{k_x}^{(0)} - 2KS - 2JS\right] \psi_1^A, \label{boundary_lower}
\\
 - 2JS \cos\left(\frac{1}{2}k_x\delta_x\right) \psi_{2N}^A &= \left[\epsilon_{k_x}^{(0)} - 2KS - 2JS\right] \psi_{2N}^B, \label{boundary_upper}
\end{align}
\end{subequations}
where the former (latter) is associated with the lower (upper) edge. Making use of the ansatz [Eq.~(\ref{ansatz})] and the Harper's equation [Eq.~(\ref{harper_A})], the boundary condition of the lower edge [Eq.~(\ref{boundary_lower})] leads to 
\begin{equation} \label{parameter}
z=-\frac{\epsilon_{k_x}^{(0)} - 2KS - 2JS}{2JS\cos(\frac{1}{2}k_x\delta_x)},
\end{equation}
which, when plugged into Eq.~(\ref{d_ribbon}), results in $z=\pm1$. Thus, the dispersions of the chiral bulk-edge magnon bands, to the zeroth order in $J'$ and $D$, read
\begin{subequations} \label{spec_edge0}
\begin{align}
\epsilon_{k_x,R}^{(0)} &= 2KS+2JS - 2JS \cos \left(\frac{1}{2} k_x\delta_x \right), \label{spec_edge0_A}
\\
\epsilon_{k_x,L}^{(0)} &=2KS+ 2JS + 2JS \cos \left(\frac{1}{2} k_x\delta_x\right),\label{spec_edge0_B}
\end{align}
\end{subequations}
which are labeled as Eq.~(\textcolor{blue}4) in the main text. We mention that the resolved $z=\pm1$ reflects the bulk-edge nature of the chiral magnon bands. For $z=\pm1$, the normalized wave functions deriving from Eq.~(\ref{ansatz}) read
\begin{subequations} \label{wf}
\begin{align}
\ket{\Psi_R} &= \frac{1}{\sqrt {4N}}(\overbrace{1, 1, \cdots, 1, 1}^{2N}, \overbrace{1, 1, \cdots, 1, 1}^{2N})^T, \label{wf_A}
\\
\ket{\Psi_L} &= \frac{1}{\sqrt{4N}} (\underbrace{1, -1, \cdots, 1, -1}_{2N}, \underbrace{-1, 1, \cdots, -1, 1}_{2N})^T. \label{wf_B}
\end{align}
\end{subequations}
These wave functions are uniformly distributed across the ribbon [red dotted lines, Figs.~\ref{figs2}(b) and~\ref{figs2}(c)]. Although we obtain the dispersions [Eq.~(\ref{spec_edge0})] by only considering the boundary condition of the lower edge [Eq.~(\ref{boundary_lower})], the spatial homogeneity of the bulk-edge magnon bands guarantees that the boundary condition of the upper edge [Eq.~(\ref{boundary_upper})] is also satisfied. The effect of the next-nearest-neighbor interactions $J'$ is incorporated perturbatively as
\begin{subequations}
\begin{align}
\bra{\Psi_R} \mathcal H_{k_x}^{J'} \ket{\Psi_R} =& -2J'S\cos(k_x\delta_x)-4J'S\cos\left(\frac{1}{2}k_x\delta_x\right) +6J'S-\frac{4}{N}J'S\sin^2\left(\frac{1}{4}k_x\delta_x\right),
\\
\bra{\Psi_L} \mathcal H_{k_x}^{J'} \ket{\Psi_L} =& -2J'S\cos(k_x\delta_x)+4J'S\cos\left(\frac{1}{2}k_x\delta_x \right) +6J'S-\frac{4}{N}J'S\cos^2\left(\frac{1}{4}k_x\delta_x\right),
\end{align}
\end{subequations}
where $\mathcal H_{k_x}^{J'} = \mathcal H_{k_x}|_{K=J=D=0}$. Therefore, the dispersions of the chiral bulk-edge magnon bands to $O(J')$ are
\begin{subequations} \label{spec_edge}
\begin{align}
\epsilon_{k_x,R} &\approx  2KS+2JS - 2JS \cos \left(\frac{1}{2} k_x\delta_x\right) + 6J'S -2J'S\cos(k_x\delta_x)-4J'S\cos\left(\frac{1}{2}k_x\delta_x\right), \label{spec_edge_a}
\\
\epsilon_{k_x,L} &\approx 2KS+ 2JS + 2JS \cos \left(\frac{1}{2} k_x\delta_x \right)+6J'S -2J'S\cos(k_x\delta_x)+4J'S\cos \left(\frac{1}{2}k_x\delta_x \right), \label{spec_edge_b}
\end{align}
\end{subequations}
where we take $N\gg 1$ and neglect the $O(1/N)$ terms. 

We now analyze the chiral edge magnon bands in the presence of the Dzyaloshinskii-Moriya interaction. Interestingly, at the level of the first-order perturbation, we find $\bra{\Psi_R} \mathcal H_{k_x}^{D} \ket{\Psi_R}=\bra{\Psi_L} \mathcal H_{k_x}^{D} \ket{\Psi_L}=0$, where $\mathcal H_{k_x}^{D}=\mathcal H_{k_x}|_{K=J=J'=0}$, implying that the Dzyaloshinskii-Moriya interaction does not introduce corrections of order $O(D)$. Therefore, Eq.~(\ref{spec_edge}) should be sufficient to analytically characterize the dispersions of the chiral edge magnon bands. A more accurate characterization requires the incorporation of the Dzyaloshinskii-Moriya interaction into the Harper's equations [Eq.~(\ref{harper})] and boundary conditions [Eq.~(\ref{boundary})], which must then be solved self-consistently even in the absence of the next-nearest-neighbor exchange interactions and anisotropy \cite{pantaleon2017, pantaleon2018}, and thus is less useful for an analytic discussion.

\section{Design lattice deformation for the uniaxial strain}
\label{s4}
In the main text, we mention that the uniaxial strain $\epsilon_{yy}=\lambda y$ may arise from an out-of-plane lattice deformation $h(\bm r)$. We here derive the explicit from of $h(\bm r)$. Without loss of generality, we take the strain parameter $\lambda>0$ and place the honeycomb ferromagnet in the $y\geq 0$ space.

We consider two neighboring lattice sites in a strain-free honeycomb ferromagnet located at $\bm r$ and $\bm r+\bm \rho$. According to Ref.~\cite{lantagne2020}, the uniaxial strain changes the spacing between the two sites to
\begin{equation} \label{S1}
S=|\bm \rho|+\frac{\bm\rho \cdot \bm \epsilon \cdot \bm \rho}{|\bm \rho|} = \rho+ \frac{\rho_y^2}{\rho} \epsilon_{yy},
\end{equation}
where $\bm \epsilon=\epsilon_{yy} \hat y \hat y$ and $\rho_y=\bm\rho\cdot \hat y$. In the presence of the out-of-plane lattice deformation $h(\bm r)$, the two aforementioned lattice sites are relocated to $\bm r+h(\bm r)\hat z$ and $\bm r+\bm \rho+ h(\bm r+\bm \rho)\hat z$, respectively. Their spacing then becomes
\begin{equation} \label{S2}
\begin{split}
S&=|\bm \rho + h(\bm r+\bm \rho)\hat z- h(\bm r)\hat z|
\approx |\bm \rho + \hat z \bm \rho \cdot\bm\nabla h| =\sqrt{\rho^2 + (\bm \rho \cdot\bm\nabla h)^2} \approx \rho+ \frac{(\bm \rho \cdot\bm\nabla h)^2}{2\rho}.
\end{split}
\end{equation}
A comparison between Eqs.~(\ref{S1}) and~(\ref{S2}) leads to 
\begin{equation}
\partial_x h =0, \qquad \partial_yh=\sqrt{2\epsilon_{yy}},
\end{equation}
from which the required out-of-plane lattice deformation can be solved as
\begin{equation}
h(\bm r)=\frac{2}{3}\sqrt{2\lambda} y^{\frac{3}{2}}+C.
\end{equation}
Since the integration constant represents a global shift, we set $C=0$ for simplicity.

\section{Chiral magnon pseudo-Landau levels in strained zigzag ribbons}
\label{s5}
In the main text, we have derived the chiral magnon pseudo-Landau levels induced by a uniaxial strain $\epsilon_{yy}=\lambda y$ with a nearest-neighbor model. In this section, we will incorporate the effects of next-nearest-neighbor interactions and show that such effects do not qualitatively alter the dispersions of the strain-induced magnon pseudo-Landau levels.

In the zigzag ribbon, the uniaxial strain preserves the $x$-direction translational symmetry but causes an out-of-plane bend $h(x,y)=\sqrt{8\lambda y^3/9}$ [Fig.~\textcolor{blue}{1}(e) of the main text], which spatially modulates the lattice site spacing and thus the interactions $J_j$, $J'_j$ and $D_j$. The spatial modulation is assumed to be exponential as \cite{liu2021a}
\begin{subequations} \label{sub}
\begin{align}
J_j(\bm r) &= J \exp \left[-\gamma\frac{\tilde\alpha_j(\bm r)- \alpha_j}{ \alpha_j} \right] = J \exp \left\{-\gamma \frac{[\bm \alpha_j \cdot \bm \nabla h(\bm r)]^2}{2\alpha_j^2} \right\} , \label{sub_A}
\\
J'_j(\bm r) &= J' \exp \left[-\gamma\frac{\tilde\beta_j(\bm r)- \beta_j}{\beta_j} \right] = J' \exp \left\{-\gamma \frac{[\bm \beta_j \cdot \bm \nabla h(\bm r)]^2}{2\beta_j^2} \right\} , \label{sub_B}
\\
D_j(\bm r) &= D \exp \left[-\gamma\frac{\tilde\beta_j(\bm r)- \beta_j}{\beta_j} \right] = D \exp \left\{-\gamma \frac{[\bm \beta_j \cdot \bm \nabla h(\bm r)]^2}{2\beta_j^2} \right\} , \label{sub_C}
\end{align}
\end{subequations}
where $\gamma \simeq 1$ is the Gr\"uneisen parameter for the honeycomb ferromagnet, $j=1,2,3$ is the index for nearest and next-nearest neighbors, $\alpha_j=a$ and $\beta_j=\sqrt 3a$ are the magnitudes of the strain-free nearest-neighbor and nearest-nearest-neighbor vectors, and $\tilde \alpha_j (\bm r)= \alpha_j + [\bm \alpha_j \cdot \bm \nabla h(\bm r)]^2/(2\alpha_j)$ and $\tilde \beta_j (\bm r)= \beta_j + [\bm \beta_j \cdot \bm \nabla h(\bm r)]^2/(2\beta_j)$ are the magnitudes of the strain-modulated nearest-neighbor and next-nearest-neighbor vectors. The forms of $\tilde \alpha_j (\bm r)$ and $\tilde \beta_j(\bm r)$ are determined according to Eq.~(\ref{S2}). Equation~(\ref{sub}) shows the form of Eq.~(\textcolor{blue}5) of the main text. Note that the easy-axis anisotropy is invariant under strain because of its onsite form.

For a strained honeycomb ferromagnet with $\lambda a\ll 1$, the spin wave dispersion is governed by the Heisenberg Hamiltonian [Eq.~(\ref{H_periodic})] with anisotropic in-plane exchange and Dzyaloshinskii-Moriya interactions [Eq.~(\ref{sub})]. Upon the Holstein-Primakoff transformation, the magnon tight-binding Hamiltonian under strain reads
\begin{equation} \label{tb_periodic_strain}
\begin{split}
\tilde H =& \sum_{\bm r,j} \left[ \frac{1}{3}g\mu_BB_z(\bm r)+J_j(\bm r)S+2J'_j(\bm r)S \right](a_{\bm r}^\dagger a_{\bm r} +b_{\bm r}^\dagger b_{\bm r}) +\sum_{\bm r} 2K S (a_{\bm r}^\dagger a_{\bm r} +b_{\bm r}^\dagger b_{\bm r} )- \sum_{\bm r, j} \left[ J_j(\bm r)S a_{\bm r}^\dagger b_{\bm r+\bm \alpha_j-\bm \alpha_1}\right. 
\\
&\left. + J'_j(\bm r)S a_{\bm r}^\dagger a_{\bm r+\bm \beta_j} + J'_j(\bm r)S b_{\bm r}^\dagger b_{\bm r+\bm \beta_j} - iD_j(\bm r)Sa_{\bm r}^\dagger a_{\bm r+\bm \beta_j} +iD_j(\bm r)S b_{\bm r}^\dagger b_{\bm r+\bm \beta_j} +\text{H.c.} \right],
\end{split}
\end{equation}
where we again ignore the ferromagnetic ground state energy but introduce a Zeeman field $B_z(\bm r)$ coupled to the magnetic dipole moment $g\mu_B$ of a magnon, with $g$ being the gyromagnetic ratio and $\mu_B$ being the Bohr magneton. The Zeeman field serves as an extra tuning knob for the onsite effect. In particular, the onsite energy in Eq.~(\ref{tb_periodic_strain}) can be restored to $\mathcal K=2KS+3JS+6J'S$ [i.e., the onsite energy in Eq.~(\ref{tb_periodic})] once the Zeeman field is tuned to
\begin{equation} \label{Bz}
B_z(\bm r)=-\frac{S}{g\mu_B} \sum_j \left[ \delta J_j(\bm r)+2\delta J'_j(\bm r) \right],
\end{equation}
where $\delta J_j(\bm r) = J_j(\bm r)-J$ and $\delta J'_j(\bm r) = J'_j(\bm r)-J'$. For simplicity, such a Zeeman field is adopted in both the main text and this supplemental material.  

The preservation of the $x$-direction translational symmetry by the applied uniaxial strain allows the application of the partial Fourier transform $(a_{\bm r}, b_{\bm r})^T=\tilde N_{\text{uc}}^{-1/2}\sum_{k_x} e^{ik_xx}(a_{k_x,y}, e^{ik_x\delta_x/2}b_{k_x,y})^T$, where $\tilde N_{\text{uc}}$ is the number of ribbon unit cells [Fig.~\textcolor{blue}{1}(e) of the main text]. The resulting ribbon tight-binding Hamiltonian reads
\begin{equation} \label{tb_ribbon_strain}
\begin{split}
\tilde H&=\sum_{k_x,y}\mathcal K \left( a_{k_x,y}^\dagger a_{k_x,y} + b_{k_x,y}^\dagger b_{k_x,y} \right) -\sum_{k_x,y}  b_{k_x,y}^
\dagger \left[ 2J_1(y)\cos\left(\frac{1}{2}k_x\delta_x\right)+J_3(y) \hat s_{\delta_y} \right]S a_{k_x,y}  
\\
&-\sum_{k_x,y}  a_{k_x,y}^
\dagger \left[ 2J_1(y)\cos\left(\frac{1}{2}k_x\delta_x\right)+J_3(y) \hat s_{-\delta_y} \right]S b_{k_x,y} +\sum_{k_x,y} \left( a_{k_x,y}^\dagger \hat{\mathcal T}_{k_x,y} a_{k_x,y}+ b_{k_x,y}^\dagger \hat{\mathcal T}_{k_x,y} b_{k_x,y} \right) 
\\ 
&+\sum_{k_x,y} \left( a_{k_x,y}^\dagger \hat{\mathcal D}_{k_x,y} a_{k_x,y}- b_{k_x,y}^\dagger \hat{\mathcal D}_{k_x,y} b_{k_x,y} \right),
\end{split}
\end{equation}
where the shift operator $\hat s_{\pm \delta_y}$ satisfies $\hat s_{\pm \delta_y}\mu_{k_x,y}=\mu_{k_x,y \pm \delta_y}$ with the sublattice index $\mu=a,b$, and for transparency we also define the following two associated operators $\hat{\mathcal T}_{k_x,y} =-2J_1'(y) S \cos(\tfrac{1}{2}k_x\delta_x) ( \hat s_{\delta_y}+ \hat s_{-\delta_y}) - 2J'_3(y)S\cos(k_x\delta_x)$ and $\hat{\mathcal D}_{k_x,y} = 2 D_3(y)S\sin(k_x\delta_x) -2D_1(y) S \sin(\tfrac{1}{2}k_x\delta_x) ( \hat s_{\delta_y}+ \hat s_{-\delta_y})$. When we set $J'=D=0$, Eq.~(\ref{tb_ribbon_strain}) is reduced to Eq.~(\textcolor{blue}6) of the main text. In the continuum limit (i.e., the lattice constant $a\rightarrow 0$), the shift operator is reduced in the linear order to $\hat s_{\pm \delta_y}=1\pm \delta_y\partial_y$. The two associated operators are then reduced to scalars as $\mathcal T_{k_x,y}=-4J_1'(y) S \cos(\tfrac{1}{2}k_x\delta_x) - 2J'_3(y)S\cos(k_x\delta_x)$ and $\mathcal D_{k_x,y} = -4D_1(y) S \sin(\tfrac{1}{2}k_x\delta_x) + 2 D_3(y)S\sin(k_x\delta_x)$. In the sublattice basis $\psi_{k_x,y}=(a_{k_x,y}, b_{k_x,y})^T$, the ribbon tight-binding Hamiltonian [Eq.~(\ref{tb_ribbon_strain})] can be rewritten as $\tilde H=\sum_{k_x,y}\psi_{k_x,y}^\dagger \tilde{\mathcal H}_{k_x,y} \psi_{k_x,y}$ with the ribbon Bloch Hamiltonian
\begin{equation} \label{Bloch_ribbon_strain}
\tilde{\mathcal H}_{k_x,y} = -\left[2J_1(y)\cos\left(\frac{1}{2}k_x\delta_x\right)+J_3(y)\right]S \sigma^x + iJ_3(y)S \delta_y \partial_y \sigma^y +(\mathcal K + \mathcal T_{k_x,y}) \sigma^0 + \mathcal D_{k_x,y} \sigma^z,
\end{equation}
where $\sigma^{x,y,z}$ and $\sigma^0$ are again the Pauli matrices and the identity matrix. It is worth noting that near the $x$-direction projected Brillouin zone corners $k_x^\eta=\eta\tfrac{4\pi}{3\delta_x}$ ($\eta=\pm 1$), the $\sigma^x$ and $\sigma^y$ terms of Eq.~(\ref{Bloch_ribbon_strain}) constitute a standard Dirac Hamiltonian upon linearization $k_x-k_x^\eta \rightarrow -i\partial_x$, the $\sigma^0$ term is a $k_x$-dependent shift $\mathcal K + \mathcal T_{k_x,y}$, and the $\sigma^z$ term is a $k_x$-dependent mass $\mathcal D_{k_x,y}$. We thus expect strain to induce chiral magnon pseudo-Landau levels in the bulk of the zigzag ribbon.

\begin{figure*} [t]
\includegraphics[width = 17.2cm]{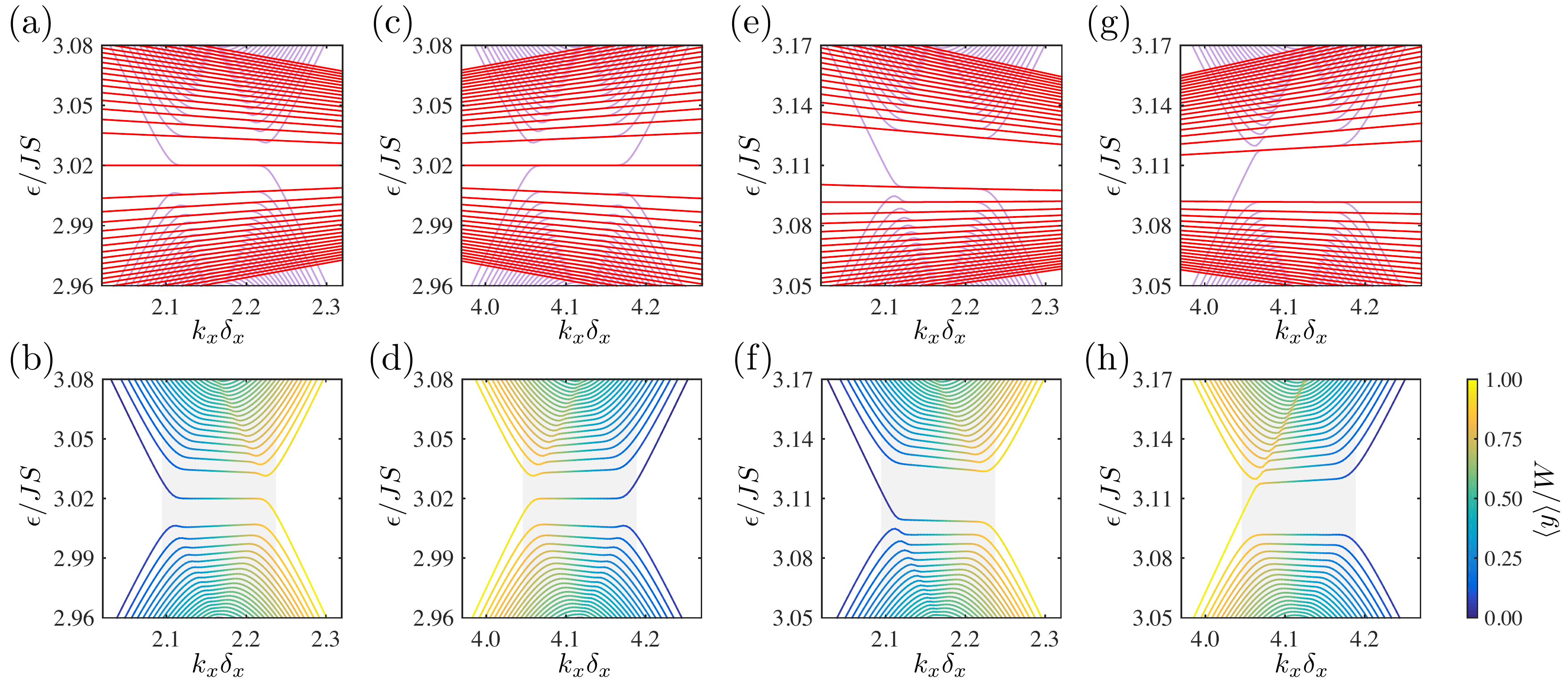}
\caption{Magnon pseudo-Landau levels in a uniaxially strained zigzag ribbon of honeycomb ferromagnets. (a, c) Magnon band structure (purple) of a strained ribbon with only nearest-neighbor interaction and easy-axis anisotropy. The red curves are the analytically predicted dispersions of magnon pseudo-Landau levels [Eq.~(\ref{pLL1})]. (b, d) Magnon band structure in panels (a) and (c), color-coded by the expectation of the position operator, i.e., $\braket y=\bra{\Psi_{k_x,y}}y\ket{\Psi_{k_x,y}}$. The shades at $[\tfrac{2\pi}{3\delta_x}, \tfrac{2\pi}{\delta_x}-\mathcal k_{\max}]$ and $[\mathcal k_{\max},\tfrac{4\pi}{3\delta_x}]$ mark the allowed ranges of the zeroth pseudo-Landau level in the momentum space. (e)--(h) Same as panels (a)--(d), except that the next-nearest-neighbor interactions are included, and the predicted pseudo-Landau level dispersions are given by Eq.~(\ref{pLL2}). We set the parameters $N=600$, $\gamma\lambda a=10^{-4}$, $K=0.01J$, and in panels (e)--(h) $J'=0.01J$ and $D=0.002J$.
} \label{figs3}
\end{figure*}

In the rest of this section, we resolve the dispersions of the chiral magnon pseudo-Landau levels. First, we discuss the $J'=D=0$ case with more details. As analyzed in the main text, the strained ribbon realizes a Su-Schrieffer-Heeger model \cite{su1979} of magnons with a domain wall at $y_0=-\frac{4}{3\gamma\lambda} \ln |-2\cos(\tfrac{1}{2}k_x\delta_x)|$. For the lattice momentum $k_x \in [0,\tfrac{2\pi}{\delta_x}]$, the domain wall can only be located inside the ribbon for the momenta $k_x\in[\tfrac{2\pi}{3\delta_x}, \tfrac{2\pi}{\delta_x}-\mathcal k_{\max}]$ and $k_x\in[\mathcal k_{\max},\tfrac{4\pi}{3\delta_x}]$, where ${\mathcal k}_{\max}=\tfrac{2}{\delta_x}\arccos[-\tfrac{1}{2}\exp(-\tfrac{3}{4}\gamma \lambda W)]$. Below the domain wall, i.e., $y<y_0$, the intercell hopping dominates and thus renders the lower sector of the ribbon unit cell topological, while the upper sector above the domain wall is topologically trivial. Since the lower sector realizes a topological Su-Schrieffer-Heeger model, an edge mode and a domain wall mode are expected \cite{su1979}. Because of the edge modulation reflected in $\bm \Lambda_A$, the edge modes at different $k_x$'s in general constitute an edge band with some complicated dispersion. In contrast, the domain wall mode is not subjected to edge modulation, being generally far away from both edges. Therefore, such domain wall modes constitute flat magnon bulk bands [Figs.~\ref{figs3}(a)--\ref{figs3}(d)] at the energy $\mathcal K_0=\mathcal K_{J'=D=0}=2KS+3JS$ for the momenta $k_x \in [\tfrac{2\pi}{3\delta_x}, \tfrac{2\pi}{\delta_x}-\mathcal k_{\max}]$ and $k_x \in [\mathcal k_{\max},\tfrac{4\pi}{3\delta_x}]$, as illustrated in Figs.~\ref{figs3}(b) and~\ref{figs3}(d).

The most important information extracted by the above band topology analysis is $y_0$, where the Su-Schrieffer-Heeger domain wall state is localized. We will now show $y_0$ is also the guiding center for the strain-induced magnon pseudo-Landau levels; indeed, the domain wall modes form the zeroth pseudo-Landau levels, which happen to be dispersionless in the absence of next-nearest-neighbor effects. For the purpose of resolving the magnon pseudo-Landau levels, it would be sufficient to study the ribbon Bloch Hamiltonian [Eq.~(\ref{Bloch_ribbon_strain})] in the vicinity of $y_0$ as
\begin{equation} \label{Dirac}
\tilde{\mathcal H}_{k_x,y}= \frac{3}{4}JS\gamma\lambda\left|-2\cos\left(\frac{1}{2}k_x\delta_x\right)\right|^{\frac{4}{3}}(y-y_0)\sigma^x +\frac{3}{2}iJSa \left|-2\cos\left(\frac{1}{2}k_x\delta_x\right)\right|^{\frac{4}{3}} \partial_y\sigma^y+\mathcal K \sigma^0,
\end{equation}
which is labeled as Eq.~(\textcolor{blue}8) in the main text. To solve the spectrum, it is more convenient to rewrite Dirac Hamiltonian [Eq.~(\ref{Dirac})] in a matrix form as
\begin{equation} \label{Dirac_mat1}
\tilde{\mathcal H}_{k_x,y}=
\begin{bmatrix}
\mathcal K_0 & \epsilon_B\hat a
\\
\epsilon_B \hat a^\dagger & \mathcal K_0
\end{bmatrix},
\end{equation}
where $\epsilon_B=\tfrac{3}{2}JS\sqrt{\gamma\lambda a}|-2\cos(\tfrac{1}{2}k_x\delta_x)|^{4/3}$ is the energy scale, and $\hat a=\tfrac{1}{\sqrt 2}(\xi+\partial_\xi)$ and $\hat a^\dagger=\tfrac{1}{\sqrt 2}(\xi-\partial_\xi)$ are the ladder operators with dimensionless coordinate $\xi=(y-y_0)/\ell_B$ and pseudomagnetic length $\ell_B=(2a/\gamma\lambda)^{1/2}$. The eigenstates of the Dirac Hamiltonian can be written as $\ket{\psi_{\pm,n>0}}=(\pm \ket{n-1}/{\sqrt 2},\ket n/{\sqrt 2})^T$ and $\ket{\psi_0}=(0,\ket 0)^T$, where $\ket n=(2^n{\sqrt \pi} n!)^{-1/2}\exp(-\xi^2/2) H_n(\xi)$ is the eigenstate of the number operator $\hat a^\dagger \hat a$, $H_n(\cdot)$ being the $n$th Hermite polynomial. The form of $\ket {\psi_0}$ indicates that the zeroth pseudo-Landau level resides on the $B$ sublattice only. This sublattice polarization is consistent with the nontrivial topology of the lower segment of the ribbon unit cell. To see this, we note that the end modes of a topological Su-Schrieffer-Heeger model exist on different sublattices of polyacetylene \cite{su1979}. The lower segment of the zigzag unit cell terminates at the lower edge on the $a_1$ site; thus the domain wall mode must appear in the bulk on a lattice site belonging in the $B$ sublattice. The eigenenergies corresponding to $\ket{\psi_{\pm,n>0}}$ and $\ket {\psi_0}$ respectively read $\epsilon_{\pm,n>0}=\mathcal K_0\pm \epsilon_B\sqrt{n}$ and $\epsilon_0=\mathcal K_0$. Explicitly, the magnon pseudo-Landau level dispersions are
\begin{equation} \label{pLL1}
\epsilon_{\nu,k_x}=2KS+ 3JS+\frac{3}{2}\text{sgn}(\nu)JS\sqrt{|\nu|\gamma\lambda a}\left|2\cos\left(\frac{1}{2}k_x\delta_x\right) \right|^{\frac{4}{3}},
\end{equation}
which is labeled as Eq.~(\textcolor{blue}9) in the main text. We emphasize that Eq.~(\ref{pLL1}) is only expected to match the bulk magnon bands accurately. The real-space position of a magnon energy band can be extracted by the expected value of the position operator, i.e., $\braket y=\bra{\Psi_{k_x}(y)} y \ket{\Psi_{k_x}(y)}$, where $\ket{\Psi_{k_x}(y)}$ is the eigenvector of the chosen magnon band [Figs.~\ref{figs3}(b) and~\ref{figs3}(d)]. Inspecting the $\nu$ and $k_x$ dependence of $\braket y$, we find that higher magnon pseudo-Landau levels (with larger $|\nu|$) are narrower, i.e., have a smaller range of $k_x$ where Eq.~(\ref{pLL1}) is valid. This is because the wave function of a higher magnon pseudo-Landau level is less localized in the real space, which may be inferred from the number of nodes of the Hermite polynomial $H_{|\nu|}(\cdot)$, and thus stays away from the edges for a smaller range of $k_x$.

We now analyze the more generic next-nearest-neighbor model. Around the domain wall, at the lowest order, recall that the next-nearest-neighbor exchange interaction produces a shift $\mathcal T_{k_x, y_0}=-4J'_1(y_0)S\cos(\tfrac{1}{2}k_x\delta_x)-2J'_3(y_0)S\cos(k_x\delta_x)$, and the next-nearest-neighbor Dyazloshinskii-Moriya interaction leads to a mass $\mathcal D_{k_x,y_0} = -4D_1(y_0)S\sin(\tfrac{1}{2}k_x\delta_x)+2D_3(y_0)S\sin(k_x\delta_x)$. The resulting Hamiltonian matrix then becomes 
\begin{equation} \label{Dirac_mat2}
\tilde{\mathcal H}_{k_x,y}=
\begin{bmatrix}
\mathcal K+\mathcal T_{k_x,y_0} + \mathcal D_{k_x,y_0} & \epsilon_B\hat a
\\
\epsilon_B \hat a^\dagger & \mathcal K+\mathcal T_{k_x,y_0} - \mathcal D_{k_x,y_0}
\end{bmatrix},
\end{equation}
whose eigenvectors are $\ket{\psi_{\pm,n>0}}=(\zeta_{\pm n}^A\ket{n-1}, \zeta_{\pm n}^B\ket{n})^T$ with $\zeta_{\pm n}^A=\pm[\tfrac{1}{2}\pm\tfrac{1}{2} (n\epsilon_B^2/\mathcal D_{k_x,y_0}^2+1)^{-1/2}]^{1/2}$ and $\zeta_{\pm n}^B=[\tfrac{1}{2}\mp \tfrac{1}{2} (n\epsilon_B^2/\mathcal D_{k_x,y_0}^2+1)^{-1/2}]^{1/2}$ and $\ket{\psi_0}=(0,\ket 0)^T$. These eigenvectors are respectively associated with the eigenenergies $\epsilon_{\pm,n>0}=\mathcal K + \mathcal T_{k_x,y_0} \pm (n\epsilon_B^2+\mathcal D_{k_x,y_0}^2)^{1/2}$ and $\epsilon_0=\mathcal K + \mathcal T_{k_x,y_0}-\mathcal D_{k_x,y_0}$. In a more compact form, the magnon pseudo-Landau level dispersions read
\begin{equation} \label{pLL2}
\epsilon_{\nu,k_x}=\mathcal K + \mathcal T_{k_x,y_0}+ \text{sgn}(\nu)\sqrt{|\nu|\epsilon_B^2+\mathcal D_{k_x,y_0}^2}-\delta_{\nu,0} \mathcal D_{k_x,y_0},
\end{equation}
where $\delta_{i,j}$ is the Kronecker delta. Equation~(\ref{pLL2}) reveals two effects resulting from the next-nearest-neighbor interactions. On the one hand, the shift $\mathcal T_{k_x,y_0}$ associated with the exchange interaction tilts the magnon energy bands. At the two valleys [Figs.~\ref{figs3}(e) and~\ref{figs3}(g)], the magnon energy bands are tilted oppositely such that the magnon pseudo-Landau levels with $\nu\geq 0$ ($\nu<0$) become steeper (flatter). On the other hand, the Haldane mass $\mathcal D_{k_x,y_0}$ arising from the Dzyaloshinskii-Moriya interaction gives rise to a topologically nontrivial band gap. Consequently, the zeroth magnon pseudo-Landau levels at the two valleys are pushed in different directions [Figs.~\ref{figs3}(e) and~\ref{figs3}(g)]. Moreover, a pair of counter-propagating edges [Figs.~\ref{figs3}(f) and~\ref{figs3}(h)] emerge in the gap as required by the bulk-boundary correspondence.

\section{Non-conservation of magnons in chiral anomaly}
\label{s6}
In the main text, we have derived the anomaly equation [Eq.~(\textcolor{blue}{12}) of the main text] of the (1+1)-dimensional magnon chiral anomaly. This magnon chiral anomaly results from the pumping of magnons by the magnetic field gradient $\partial_xB$ and/or the temperature gradient $\partial_x T$ according to the semiclassical equation of motion
\begin{equation} \label{eom}
\hbar \frac{dk_x}{dt} = g\mu_B\partial_xB -\frac{\epsilon_{k_x}}{T}\partial_x T,
\end{equation}
which is labeled as Eq.~(\textcolor{blue}{10}) in the main text. Equation~(\ref{eom}) incorporates two driving forces: (i) the Zeeman force and (ii) the entropic force  \cite{huebener2001}. In this section, we show that magnons exhibit non-conservation after pumped by these forces.

We write down the magnon concentration variations for the right-moving and left-moving chiral magnon bands as
\begin{subequations} \label{nRL}
\begin{align} 
n_R &=\int_{\epsilon_0+\epsilon_5}^{\epsilon_0} \rho(\epsilon) n_B(\epsilon) d\epsilon \approx \rho(\epsilon_0)  \int_{\epsilon_0+\epsilon_5}^{\epsilon_0} n_B(\epsilon) d\epsilon = -\frac{1}{hv_R(\epsilon_0)} \sum_{j=0} \frac{n_B^{(j)}(\epsilon_0)}{(j+1)!} \epsilon_5^{j+1} \approx- \frac{n_B(\epsilon_0)}{2\pi \hbar v_R(\epsilon_0)}\epsilon_5, \label{nR}
\\
n_L &=\int_{\epsilon_0-\epsilon_5}^{\epsilon_0} \rho(\epsilon) n_B(\epsilon) d\epsilon \approx \rho(\epsilon_0) \int_{\epsilon_0-\epsilon_5}^{\epsilon_0}  n_B(\epsilon) d\epsilon = -\frac{1}{hv_R(\epsilon_0)} \sum_{j=0} \frac{n_B^{(j)}(\epsilon_0)}{(j+1)!} (-\epsilon_5)^{j+1} \approx \frac{n_B(\epsilon_0)}{2\pi \hbar v_R(\epsilon_0)}\epsilon_5, \label{nL}
\end{align}
\end{subequations}
where $n_B(\epsilon)=(e^{\epsilon/k_BT}-1)^{-1}$ is the Bose-Einstein distribution function, $\rho(\epsilon)=1/hv_R(\epsilon)$ is the density of states for the one-dimensional chiral bands with $v_R(\epsilon)$ being the velocity of the right-moving chiral band, and $\epsilon_0$ is the magnon population edge when the two magnon reservoirs are in equilibrium [Fig.~\textcolor{blue}4(b) of the main text]. For the magnon population edge in consideration $\epsilon_0\simeq 3JS$, the chiral magnon bands exhibit almost linear dependence on $k_x$, making $v_R(\epsilon)$ approximately a constant, giving rise to the first approximation in Eq.~(\ref{nRL}). The second approximation in Eq.~(\ref{nRL}) is made in the limit $|\epsilon_5|\ll k_BT \ll \epsilon_0$. The combination of Eqs.~(\ref{nR}) and~(\ref{nL})  is written as Eq.~(\textcolor{blue}{11}) in the main text. To the linear order of $\epsilon_5$, we have $n_R+n_L\approx 0$, meaning that the number of magnons pumped into (out of) the right-moving chiral band is approximately the same as that of the magnons pumped out of (into) the left-moving chiral band, as illustrated in Fig.~\textcolor{blue}4(c) [Fig.~\textcolor{blue}4(d)]. Beyond the linear order, we find 
\begin{equation}
n_R+n_L \approx -\frac{1}{hv_R(\epsilon_0)} \sum_{j=0} \frac{n_B^{(j)}(\epsilon_0)}{(j+1)!} \left[\epsilon_5^{j+1}+(-\epsilon_5)^{j+1}\right] =\frac{k_BT}{hv_R(\epsilon_0)}\ln \left[ \frac{e^{2\epsilon_0/k_BT}-2e^{\epsilon_0/k_BT}+1}{e^{2\epsilon_0/k_BT}-2\cosh\left(\frac{\epsilon_5}{k_BT}\right)e^{\epsilon_0/k_BT}+1} \right]>0, 
\end{equation}
which indicates an increase of the total magnon number. This is because magnons are bosonic collective excitations immune to particle conservation law. From another point of view, if the magnon number were conserved, the magnon populations after pumping [Figs.~\textcolor{blue}4(c) and~\textcolor{blue}4(d) of the main text] would be of less energy than those before $\partial_x B$ and $\partial_x T$ are switched on [Figs.~\textcolor{blue}4(b) of the main text], rendering the pumping unrealistically spontaneous. In fact, energy must be acquired from $\partial_x B$ and $\partial_x T$ to pump the magnons. To see this explicitly, we calculate the energy accumulation on the right-moving and left-moving chiral bands as
\begin{subequations} \label{URL}
\begin{align}
U_R&=\int_{\epsilon_0+\epsilon_5}^{\epsilon_0} \epsilon \rho(\epsilon) n_B(\epsilon) d\epsilon \approx \rho(\epsilon_0) \int_{\epsilon_0+\epsilon_5}^{\epsilon_0} \epsilon n_B(\epsilon) d\epsilon =-\frac{1}{hv_R(\epsilon_0)}\sum_{j=0}\frac{n_B^{(j)}(\epsilon_0)}{(j+1)!}\left( \epsilon_0+ \frac{j+1}{j+2}\epsilon_5\right) \epsilon_5^{j+1},
\\
U_L &=\int_{\epsilon_0-\epsilon_5}^{\epsilon_0} \epsilon \rho(\epsilon) n_B(\epsilon) d\epsilon \approx  \rho(\epsilon_0) \int_{\epsilon_0-\epsilon_5}^{\epsilon_0} \epsilon n_B(\epsilon) d\epsilon =-\frac{1}{hv_R(\epsilon_0)}\sum_{j=0}\frac{n_B^{(j)}(\epsilon_0)}{(j+1)!}\left( \epsilon_0- \frac{j+1}{j+2}\epsilon_5\right) (-\epsilon_5)^{j+1},
\end{align}
\end{subequations}
where we again pull $\rho(\epsilon)$ out of the integrals because it is approximately a constant. The total energy acquisition from the external magnetic field and temperature field is then
\begin{equation}
\begin{split}
U_R+U_L \approx \left. -\frac{2}{hv_R(\epsilon_0)}\sum_{j=0}\frac{\epsilon_5^{2j+2}}{(2j+2)!}\frac{d^{2j+1}}{d\epsilon^{2j+1}} \epsilon n_B(\epsilon)\right|_{\epsilon_0} =2\mathcal F(\epsilon_0) - \mathcal F(\epsilon_0+\epsilon_5)- \mathcal F(\epsilon_0-\epsilon_5)>0,
\end{split}
\end{equation}
where $\mathcal F(\epsilon)=\int \epsilon n_B(\epsilon) d\epsilon$. Through integration by part (also see Sec.~\ref{s8}), we find $\mathcal F(\epsilon)= \tfrac{k_B^2T^2}{hv_R(\epsilon_0)} \{ \text{Li}_2[-n_B(\epsilon)]+\tfrac{1}{2}\ln^2[n_B(\epsilon)]-\tfrac{\epsilon^2}{2k_B^2T^2}\}$ with $\text{Li}_2(z)=\sum_{j=1} z^j/j^2$ being the dilogarithm, provided that the constant of integration is set to zero. It is also straightforward to check that $\mathcal F(\epsilon)$ is a concave function of $\epsilon$, which necessarily leads to $\mathcal F(\epsilon_0)>\tfrac{1}{2}[\mathcal F(\epsilon_0+\epsilon_5)+\mathcal F(\epsilon_0-\epsilon_5)]$. Alternatively, in the $|\epsilon_5|\ll k_BT\ll \epsilon_0$ limit, $U_R+U_L\approx -\frac{\epsilon_5^2}{hv_R(\epsilon_0)} [n_B(\epsilon_0)+\epsilon_0n_B'(\epsilon_0)]>0$. Such an increase of the total energy reflects the non-conservation of magnons in the (1+1)-dimensional magnon chiral anomaly.

\section{Boltzmann formalism for magnons}
\label{s7}
In the main text, we have adopted the magnon dipole conductivity and thermal conductivity [i.e., Eq.~(\textcolor{blue}{13}) of the main text] as the transport signatures of the (1+1)-dimensional magnon chiral anomaly. In this section, we will derive these transport signatures by constructing Boltzmann formalism for magnons.

\begin{figure} [t]
\includegraphics[width = 17.2cm]{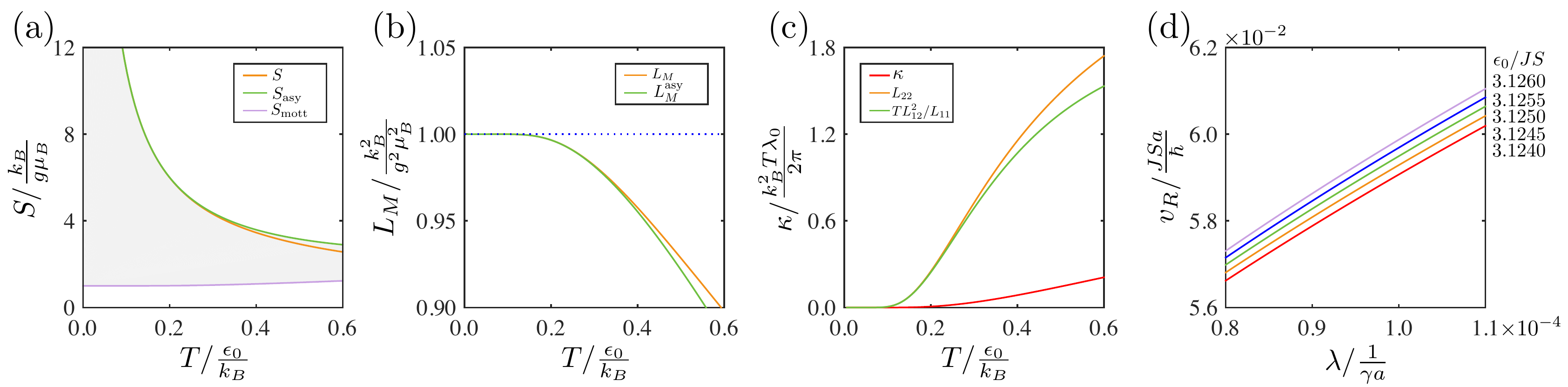}
\caption{Transport signatures of chiral magnon bands in the zigzag ribbon of honeycomb ferromagnets. (a) Magnon Seebeck coefficient. $S$ and $S_{\text{asy}}$ are respectively given by Eqs.~(\ref{seebeck1}) and~(\ref{seebeck2}). $S_{\text{mott}}$ is the first term in Eq.~(\ref{seebeck2}), and the shade marks the second term in Eq.~(\ref{seebeck2}) which is the dominant contribution scaling as $\tfrac{1}{T}$. (b) Magnon Lorenz number as a function of temperature. $L_M$ and $L_M^{\text{asy}}$ are respectively given by Eqs.~(\ref{lorenz1}) and~(\ref{lorenz2}). (c) Magnon thermal conductivity (red), calculated as the difference between $L_{22}$ (orange) and $TL_{12}^2/L_{11}$ (green). (d) Magnon drift velocity associated with the right-moving branch of the first chiral magnon pseudo-Landau level in a uniaxially strained ribbon with next-nearest-neighbor interactions. Different curves correspond to different positions of the magnon population edge. The parameter values adopted are $K/J=0.01$, $J'/J=0.01$, and $D/J=0.002$.
} \label{figs4}
\end{figure}

We first derive the nonequilibrium steady-rate magnon distribution function, labeled as $\mathcal n$. According to the semiclassical equation of motion [Eq.~(\ref{eom})], the nonequilibrium distribution function $\mathcal n$ varies at a rate
\begin{equation}
\frac{d\mathcal n}{dt}=\frac{d\mathcal n}{dk_x}\frac{dk_x}{dt}=\frac{d\mathcal n}{d\epsilon_{k_x}}v_x \left(g\mu_B\partial_xB-\frac{\epsilon_{k_x}}{T}\partial_xT \right),
\end{equation}
where the drift velocity $v_x=\tfrac{1}{\hbar}\tfrac{d\epsilon_{k_x}}{dk_x}$ is a function of $k_x$. For slowly varying magnetic and temperature fields (i.e., $\partial_xB \rightarrow 0$ and $\partial_xT \rightarrow 0$), we adopt the relaxation time approximation
\begin{equation} \label{Boltzmann}
\frac{d\mathcal n}{dt} \approx \frac{n_B-\mathcal n}{\tau},
\end{equation}
where $\tau$ is the relaxation time. We assume the major source of scattering is the intervalley one that keeps magnon concentration variations finite. We can now solve the magnon Boltzmann equation Eq.~(\ref{Boltzmann}) for the nonequilibrium distribution function. To the lowest order in external fields, approximating $d\mathcal n/d\epsilon_{k_x}\approx dn_B/d\epsilon_{k_x}$, we find
\begin{equation} \label{noneq_BE}
\mathcal n=n_B+\tau v_x \left( g\mu_B \partial_x B - \frac{\epsilon_{k_x}}{T} \partial_x T \right)  \left(-\frac{\partial n_B}{\partial \epsilon_{k_x}} \right).
\end{equation}
An alternative form of the nonequilibrium distribution function is also given in Ref.~\cite{liaobolin2014}. From Eq.~(\ref{noneq_BE}), we can construct the magnetic dipole current and the heat current carried by the magnons in the chiral magnon bands as
\begin{equation} \label{current}
\begin{split}
J_M = L_{11} \partial_x B + L_{12} (-\partial_x T),
\\
J_Q= L_{21} \partial_x B + L_{22} (-\partial_x T).
\end{split}
\end{equation}
where the transport coefficients read
\begin{subequations} \label{coeff}
\begin{align}
L_{11} &= \frac{1}{2\pi} g^2\mu_B^2 \mathcal L_0, 
\\
L_{12} &= \frac{1}{2\pi} g\mu_B \frac{1}{T} \mathcal L_1 = \frac{1}{T} L_{21},
\\
L_{22} &= \frac{1}{2\pi} \frac{1}{T} \mathcal L_2.
\end{align}
\end{subequations}
In Eq.~(\ref{coeff}), we have defined for transparency the following quantity 
\begin{equation} \label{Lm1}
\mathcal L_m = \int v_x^2 \tau \epsilon_{k_x}^m \left( -\frac{\partial n_B}{\partial \epsilon_{k_x}} \right)dk_x  = \int_{\epsilon_0}^{+\infty} \lambda_m \left( -\frac{\partial n_B}{\partial \mathcal E} \right) d\mathcal E, 
\end{equation}
where $m=0,1,2$ and the auxiliary function $\lambda_m$ explicitly reads
\begin{equation} \label{lambda1}
\lambda_m(\mathcal E) =  \mathcal E^m \lambda_0(\mathcal E)=\mathcal E^m \int dk_x\, v_x^2 \tau \delta(\mathcal E-\epsilon_{k_x}).
\end{equation}
For $\mathcal E > \epsilon_0$, the derivative $-\partial n_B/\partial \mathcal E$ in Eq.~(\ref{Lm1}) decays exponentially over the scale of $k_BT$. This implies that the main contribution to $\mathcal L_m$ comes from $\mathcal E-\epsilon_0 \lesssim k_BT$. Therefore, we expand the auxiliary function as
\begin{equation} \label{lambda2}
\lambda_m (\mathcal E) = \left.\lambda_m (\epsilon_0) + \sum_{s=1} \frac{1}{s!}\frac{d^s \lambda_m}{d\mathcal E^s}\right|_{\epsilon_0} (\mathcal E-\epsilon_0)^s.
\end{equation}
Plugging Eq.~(\ref{lambda2}) into Eq.~(\ref{Lm1}), we arrive at
\begin{equation} \label{Lm2}
\mathcal L_m = \left. \lambda_m(\epsilon_0) n_B(\epsilon_0) + \sum_{s=1} \frac{1}{s!}\frac{d^s \lambda_m}{d\mathcal E^s}\right|_{\epsilon_0} (k_BT)^s a_s,
\end{equation}
where the coefficient $a_{s>0}$ reads
\begin{equation} \label{as}
a_s = \int_{\epsilon_0}^{+\infty} \left(\frac{\mathcal E-\epsilon_0}{k_BT}\right)^s  \left( -\frac{\partial n_B}{\partial \mathcal E} \right) d\mathcal E.
\end{equation}
Since we are interested in $\mathcal E-\epsilon_0\lesssim k_BT$, the  chiral magnon bands at $\mathcal E$ exhibit almost linear dispersion $\epsilon_{k_x}\propto k_x$. Thus, the velocity $v_x$ is approximately a constant. Therefore, $\lambda_m(\mathcal E)$ acquires its $\mathcal E$ dependence almost entirely from the prefactor $\mathcal E^m$ in Eq.~(\ref{lambda1}) such that $\tfrac{d^s \lambda_m}{d\mathcal E^s}|_{\epsilon_0}\approx \tfrac{m!}{(m-s)!}\epsilon_0^{m-s}\lambda_0(\epsilon_0)$. We can thus cut off the sum in Eqs.~(\ref{lambda2}) and~(\ref{Lm2}) at $s=m$ and only need to evaluate the coefficients $a_1$ and $a_2$ 
\begin{equation} \label{a12}
\begin{split}
a_1 &= \mathcal C_1[n_B(\epsilon_0)] - \frac{\epsilon_0}{k_BT} \mathcal C_0[n_B(\epsilon_0)],
\\
a_2 &= \mathcal C_2[n_B(\epsilon_0)] - 2\frac{\epsilon_0}{k_BT} \mathcal C_1[n_B(\epsilon_0)] + \frac{\epsilon_0^2}{k_B^2T^2} \mathcal C_0[n_B(\epsilon_0)], 
\end{split}
\end{equation}
where $\mathcal C_0(z)=z$, $\mathcal C_1(z)=(z+1)\ln (z+1) - z\ln z$, and $\mathcal C_2(z)=(z+1)\ln^2(\frac{z+1}{z}) - \ln^2z-2\text{Li}_2(-z)$. Some useful integrals adopted to evaluate $a_{1,2}$ through Eq.~(\ref{as}) are listed in Sec.~\ref{s8} following the present section. Making use of Eqs.~(\ref{Lm2}) and~(\ref{a12}), we have
\begin{equation}
\mathcal L_m = \lambda_0(\epsilon_0) k_B^mT^m \mathcal C_m[n_B(\epsilon_0)],
\end{equation} 
which, upon substitution into Eq.~(\ref{coeff}), results in the following transport coefficients 
\begin{subequations}
\begin{align}
L_{11} &= \frac{g^2\mu_B^2}{2\pi } \lambda_0(\epsilon_0)\mathcal C_0[n_B(\epsilon_0)],
\\
L_{12} &= \frac{g\mu_B k_B}{2\pi }  \lambda_0(\epsilon_0) \mathcal C_1[n_B(\epsilon_0)] = \frac{1}{T}L_{21},
\\
L_{22} &=  \frac{k_B^2T}{2\pi }  \lambda_0(\epsilon_0)\mathcal C_2[n_B(\epsilon_0)].
\end{align}
\end{subequations}
The transport coefficient $L_{11}$ measures the magnon dipole current response to a magnetic field gradient in the absence of a temperature gradient. By analogy with the electrical conductivity of Bloch electrons, we refer to $L_{11}$ as the magnon dipole conductivity
\begin{equation} \label{sigma}
\sigma(\epsilon_0)=L_{11}=\frac{g^2\mu_B^2}{2\pi } \lambda_0(\epsilon_0)\mathcal C_0[n_B(\epsilon_0)].
\end{equation}
When the temperature gradient is switched on, it also contributes to the magnon dipole current by means of diffusion. In the case of a vanishing total dipole current, we can define the magnon Seebeck coefficient
\begin{equation} \label{seebeck1}
S(\epsilon_0)=\frac{L_{12}}{L_{11}}=\frac{ k_B}{g\mu_B} \frac{\mathcal C_1[n_B(\epsilon_0)]}{\mathcal C_0[n_B(\epsilon_0)]}.
\end{equation}
In the low temperature limit $T\ll\epsilon_0/k_B$, the Bose-Einstein distribution function is reduced to $n_B(\epsilon_0) \approx e^{-\epsilon_0/k_BT}\ll 1$, and we have $\mathcal C_1[n_B(\epsilon_0)]/\mathcal C_0[n_B(\epsilon_0)] \approx 1+\epsilon_0/k_BT$. Meanwhile, recalling that $\lambda_0(\mathcal E)$ barely depends on $\mathcal E$, we find $\sigma(\epsilon_0)$ to be approximately proportional to $n_B(\epsilon_0)$. Therefore, the Seebeck coefficient Eq.~(\ref{seebeck1}) asymptotically satisfies
\begin{equation} \label{seebeck2}
S(\epsilon_0) \approx -\frac{k_B^2T}{g\mu_B} \frac{d}{d\epsilon_0}\ln [\sigma(\epsilon_0)] + \frac{\epsilon_0}{g\mu_B} \frac{1}{T},
\end{equation}
where the first term $S_{\text{mott}}=-\tfrac{k_B^2T}{g\mu_B} \tfrac{d}{d\epsilon_0}\ln [\sigma(\epsilon_0)]$ resembles the Mott relation of electrons $S(\mu)=-\tfrac{\pi^2}{3}\tfrac{k_B^2T}{e}\tfrac{d}{d\mu}\ln[\sigma(\mu)]$ with $\sigma(\mu)$ being the electrical conductivity \cite{cutler1969}. Here the magnon dipole moment $g\mu_B$ is parallel to the electron charge $-e$, and the magnon population edge $\epsilon_0$ plays the role of the electron chemical potential $\mu$. However, the second term in Eq.~(\ref{seebeck2}) is in fact the dominant contribution to the Seebeck coefficient at low temperature because of the appearance of the $\tfrac{1}{T}$ factor [Fig.~\ref{figs4}(a)]. Such a violation of the Mott relation reflects the bosonic nature of magnons and implies the potential application of the honeycomb ferromagnets as ``thermo-magnetic'' devices. Keeping the condition $J_M=0$, we find that the heat current $J_Q$ is related to the temperature gradient through the magnon thermal conductivity
\begin{equation} \label{kappa}
\kappa(\epsilon_0) = L_{22}-\frac{L_{12}L_{21}}{L_{11}} =L_M T\sigma(\epsilon_0), 
\end{equation}
which follows the magnon Wiedemann-Franz law \cite{nakata2015, nakata2017} with the magnon Lorenz number 
\begin{equation} \label{lorenz1}
L_M=\frac{k_B^2}{g^2\mu_B^2}  \left\{ \frac{\mathcal C_2[n_B(\epsilon_0)]}{\mathcal C_0[n_B(\epsilon_0)]} -  \frac{\mathcal C_1^2[n_B(\epsilon_0)]}{ \mathcal C_0^2[n_B(\epsilon_0)]} \right\}.
\end{equation}
In the low temperature regime $T \ll \epsilon_0/k_B$, the magnon Lorenz number is greatly simplified to
\begin{equation} \label{lorenz2}
L_M\approx \frac{k_B^2}{g^2\mu_B^2} \left[ 1-\frac{1}{2}n_B(\epsilon_0) \right].
\end{equation}
For $T\rightarrow 0$, $n_B(\epsilon_0)$ vanishes and the Lorenz number becomes a constant $L_M=k_B^2/(g\mu_B)^2$ [Fig.~\ref{figs4}(b)]. This magnon Lorenz number exhibits a similarity to the electron Lorenz number $L_E={\pi^2k_B^2}/{(3e^2)}$ with the magnon dipole moment $g\mu_B$ playing the role of the electron charge $-e$. However, it is worth noting that the electron thermal conductivity barely depends on the thermoelectric effect at low temperature \cite{ashcroft1976}, while both $L_{22}$ and $L_{12}L_{21}/L_{11}$ in Eq.~(\ref{kappa}) have significant contributions to the magnon thermal conductivity [Fig.~\ref{figs4}(c)]. Such a difference arises from the violation of the Mott relation. In particular, the second term in Eq.~(\ref{seebeck2}) causes a pronounced thermo-magnetic effect at low temperature.

We now analyze the magnon transport associated with the (1+1)-dimensional magnon chiral anomaly in the framework of magnon Boltzmann formalism. Since $v_x$ barely depends on $k_x$, it is straightforward to find $\lambda_0(\epsilon_0)=\tfrac{2\tau}{\hbar}|v_x(\epsilon_0)|$, where $v_x(\epsilon_0)>0$ [$v_x(\epsilon_0)<0$] for the right-moving (left-moving) chiral magnon band. For transparency, we write $|v_x(\epsilon_0)|=v_R(\epsilon_0)$ in the following. Making use of Eq.~(\ref{sigma}), we find
\begin{equation} \label{transport_a}
\sigma(\epsilon_0)=\frac{g^2\mu_B^2\tau}{\pi\hbar} v_R(\epsilon_0)n_B(\epsilon_0).
\end{equation}
The chiral anomaly equation [Eq.~(\textcolor{blue}{12}) of the main text] is derived in the limit $|\epsilon_5| \ll k_BT \ll \epsilon_0$. In this limit, the magnon Lorenz number can be treated as a constant $L_M=k_B^2/(g\mu_B)^2$. Per magnon Wiedemann-Franz law, the thermal conductivity associated with the (1+1)-dimension magnon chiral anomaly reads
\begin{equation} \label{transport_b}
\kappa(\epsilon_0) =  \frac{k_B^2 T \tau}{\pi\hbar} v_R(\epsilon_0)n_B(\epsilon_0).
\end{equation}
Equations~(\ref{transport_a}) and~(\ref{transport_b}) constitute Eq.~(\textcolor{blue}{13}) of the main text. Because Eqs.~(\ref{transport_a}) and~(\ref{transport_b}) only incorporate the effects of the chiral magnon bands, the magnon population edge $\epsilon_0$ should be set inside the gap separating the chiral magnon bands from the others. The explicit evaluation of Eqs.~(\ref{transport_a}) and~(\ref{transport_b}) requires the knowledge on $v_R(\epsilon_0)$, which could be extracted from the dispersions of the chiral magnon bands derived in Secs.~\ref{s3} and~\ref{s5}.

For the unstrained ribbon discussed in Sec.~\ref{s3}, the dispersions of the chiral magnon bands are given by Eq.~(\ref{spec_edge}), regardless of the existence of the Dzyaloshinskii-Moriya interaction. We set $\epsilon_0=\mathcal K= 2KS+3JS+6J'S$ by properly tuning the attached magnon reservoirs [Fig.~\textcolor{blue}4(a) of the main text], such that the magnons begin to populate the chiral bulk-edge or edge magnon bands right in the middle of the magnon band gap in the absence or presence of the Dzyaloshinskii-Moriya interaction. The magnon drift velocity of the right-moving chiral band is then 
\begin{equation} 
v_R(\epsilon_0)=\left.\frac{1}{\hbar}\frac{d\epsilon_{k_x,R}}{dk_x}\right|_{\epsilon_0}=\frac{3JSa}{2\hbar},
\end{equation}
where $\epsilon_{k_x,R}$ is the dispersion of the right-moving chiral bulk-edge or edge magnon band [Eq.~(\ref{spec_edge_a})]. When the magnon population edge $\epsilon_0$ differs from $\mathcal K$, the form of $v_R(\epsilon_0)$ generally gets more complicated. 
 
For the uniaxially strained ribbon discussed in Sec.~\ref{s5}, without loss of generality, we tune the magnon population edge $\epsilon_0$ to intersect the first chiral magnon pseudo-Landau level at the momentum $k_0$ and the energy $\epsilon_0 = \epsilon_{\nu=1,k_x=k_0}$. In the absence of the next-nearest-neighbor effects (i.e., the case discussed in the main text), $\epsilon_{\nu,k_x}$ adopts the form of Eq.~(\ref{pLL1}). The magnon drift velocity associated with the right-moving chiral magnon pseudo-Landau level is
\begin{equation} 
\begin{split}
v_R(\epsilon_0)&=\left.\frac{1}{\hbar}\frac{d\epsilon_{1,k_x}}{dk_x}\right|_{k_0} = \frac{2\delta_x}{\hbar}JS\sqrt{\gamma\lambda a} \left|2\cos\left(\frac{1}{2}k_0\delta_x\right)\right|^{\frac 1 3} \sin\left(\frac{1}{2}k_0\delta_x\right)
\\
&=\frac{2^{\frac{5}{4}}\delta_x}{3^{\frac{7}{16}}\hbar}  (\epsilon_0-2KS-3JS)^{\frac{1}{4}}\sqrt{\left(JS\sqrt{\gamma\lambda \delta_x}\right)^{\frac 3 2} - \frac{(\epsilon_0-2KS-3JS)^{\frac 3 2}}{3^{\frac 9 8}\sqrt 2} },
\end{split}
\end{equation}
which is obviously an increasing function of the strain $\lambda$. Such a drift velocity thus causes a negative strain-resistivity for both the magnon dipole and heat. In the presence of next-nearest-neighbor interactions, $\epsilon_{\nu,k_x}$ adopts the form of Eq.~(\ref{pLL2}). The magnon drift velocity of the right-moving chiral magnon pseudo-Landau level becomes much more lengthy,
\begin{equation} \label{vR}
v_R[k_0(\epsilon_0)]= -\frac{2J'S\delta_x}{\hbar} \sin (k_0\delta_x )  + \frac{\delta_x}{\hbar} \frac{2J^2S\gamma\lambda a \left[-2\cos\left(\frac{1}{2}k_0\delta_x\right)\right]^{\frac{5}{3}} \sin\left(\frac{1}{2}k_0\delta_x\right) + 24D^2S\sin(k_0\delta_x)\cos(k_0\delta_x)}{\sqrt{16D^2\sin^2(k_0\delta_x) + J^2\gamma\lambda a \left[-2\cos\left(\frac{1}{2}k_0\delta_x\right)\right]^{\frac{8}{3}}}},
\end{equation} 
where $k_0$ is a function of both $\epsilon_0$ and $\lambda$. To extract the rather complicated $\lambda$ dependence, we fix $\epsilon_0$ in the energy window of the first chiral magnon pseudo-Landau level, then numerically solve the equation $\epsilon_0 = \epsilon_{\nu=1,k_x=k_0}$ for $k_0$ before plugging the solution into Eq.~(\ref{vR}). As shown in Fig.~\ref{figs4}(d), $v_R$ is a monotonically increasing function of $\lambda$. We thus expect the magnon dipole resistivity and thermal resistivity to decrease with an increased $\lambda$, leading to a negative strain-resistivity.

\section{Useful integrals}
\label{s8}
The evaluation of the energy accumulation $U_{R,L}$ [Eq.~(\ref{URL})] and the coefficients $a_s$ [Eq.~(\ref{as})] requires the calculation of the following integrals 
\begin{equation}
\mathcal I_m(x) = \int dx\, x^m \left(-\frac{d}{dx}\frac{1}{e^x-1}\right),
\end{equation}
where the indices $m=0, 1, 2$ are respectively associated with the magnon dipole conductivity, Seebeck coefficient, and thermal conductivity. The evaluation of $\mathcal I_0(x)$ is straightforward,
\begin{equation}
\mathcal I_0(x) = \int dx \left(-\frac{d}{dx}\frac{1}{e^x-1}\right) = - n(x),
\end{equation}
where $n(x)=(e^x-1)^{-1}$. For here and below, the constant of integration is neglected for transparency. $\mathcal I_1(x)$ can be evaluated by integrating by part as
\begin{equation}
\begin{split}
\mathcal I_1(x) &= \int dx\, x \left(-\frac{d}{dx}\frac{1}{e^x-1}\right) = - x \frac{1}{e^x-1} + \int dx \frac{e^{-x}}{1-e^{-x}} = -n(x) \ln \left[ \frac{n(x)+1}{n(x)} \right]+ \ln \left[\frac{1}{n(x)+1} \right] 
\\
&= -[n(x)+1] \ln [n(x)+1] + n(x) \ln n(x). 
\end{split}
\end{equation}
To evaluate $\mathcal I_2(x)$, we make use of the following integral
\begin{equation}
\int dz \frac{\ln (z+1)}{z}  = \sum_{j=1}^\infty \frac{(-1)^{j-1}}{j}  \int dz\, z^{j-1} = - \sum_{j=1}^\infty \frac{(-z)^j}{j^2}  = - \text{Li}_2(-z),
\end{equation}
where the Taylor expansion $\ln (z+1) = \sum_{j=1}^\infty \frac{1}{j} (-1)^{j-1} z^j$ is adopted and $\text{Li}_2(z)=\sum_{j=1}^\infty \tfrac{z^j}{j^2}$ is the dilogarithm. We evaluate $\mathcal I_2(x)$ through integration by part as
\begin{equation} 
\begin{split}
\mathcal I_2(x)  =& \int dx\, x^2 \left(-\frac{d}{dx}\frac{1}{e^x-1}\right) 
= -\frac{x^2}{e^x-1} + \int dx^2 \frac{1}{e^x-1} = -x^2 \left(\frac{1}{e^x-1}+1 \right) + \int dx^2 \frac{e^x}{e^x-1} 
\\
=& - \frac{x^2e^x}{e^x-1} -2\int dx \frac{e^x}{e^x-1} \ln \left(\frac{1}{e^x-1}\right) + 2\int dx \frac{e^x}{e^x-1} \ln \left( \frac{e^x}{e^x-1}\right),
\end{split}
\end{equation}
where we have used $x=\ln e^x = \ln( \frac{e^x}{e^x-1}) - \ln(\frac{1}{e^x-1})$ in the second line. The second term reads
\begin{equation}
\mathcal I_2^{(2)}(x) = - 2 \int  dx  \frac{e^x}{e^x-1} \ln \left(\frac{1}{e^x-1}\right) =2 \int d\left[ \ln \left(\frac{1}{e^x-1}\right)\right] \ln \left(\frac{1}{e^x-1}\right) = \left[ \ln \left(\frac{1}{e^x-1}\right)\right]^2. 
\end{equation}
The third term can be expressed in terms of the dilogarithm as
\begin{equation}
\mathcal I_2^{(3)}(x) = 2 \int dx \frac{e^x}{e^x-1} \ln \left( \frac{e^x}{e^x-1}\right) =2 \int d\left(\frac{1}{e^x-1}\right) (1-e^x) \ln \left[1+ \left(\frac{1}{e^x-1}\right) \right] = 2 \text{Li}_2\left(- \frac{1}{e^x-1}\right).
\end{equation}
To summarize, the integral $\mathcal I_2(x)$ finally reads
\begin{equation} 
\begin{split}
\mathcal I_2(x) &= -x^2 \left(\frac{1}{e^x-1}+1 \right) 
+ \left[ \ln \left(\frac{1}{e^x-1}\right)\right]^2 + 2 \text{Li}_2\left(- \frac{1}{e^x-1}\right) 
\\
&= -[n(x)+1] \ln^2 \left[\frac{n(x)+1}{n(x)}\right] + \ln^2 [n(x)] +2 \text{Li}_2[-n(x)].
\end{split}
\end{equation}

\bibliographystyle{apsrev4-1-etal-title_10authors}
\bibliography{MCAsuppl20230530}